\documentclass[twocolumn]{aastex62}

\renewcommand{\baselinestretch}{1.1}
\usepackage{textcomp}
\usepackage{gensymb}
\usepackage{epstopdf}
\usepackage{amsmath}
\usepackage{graphicx}
\usepackage{hyperref}
\usepackage{multirow}
\usepackage[nameinlink]{cleveref}
\usepackage[multiple]{footmisc}
\usepackage[normalem]{ulem}

\crefname{paragraph}{\S}{\S\S} 
\shorttitle{OGLE-2011-BLG-0950} 
\shortauthors{S. K. Terry, et al.}

\begin{document}
\pagecolor{white}
\title{\textbf{Adaptive Optics Imaging Can Break the Central Caustic Cusp Approach Degeneracy in High Magnification Microlensing Events}}

\author[0000-0002-5029-3257]{Sean K. Terry}
\affiliation{Department of Astronomy, University of California Berkeley, Berkeley, CA 94701, USA}

\author[0000-0001-8043-8413]{David P. Bennett}
\affiliation{Code 667, NASA Goddard Space Flight Center, Greenbelt, MD 20771, USA}
\affiliation{Department of Astronomy, University of Maryland, College Park, MD 20742, USA}

\author{Aparna Bhattacharya}
\affiliation{Code 667, NASA Goddard Space Flight Center, Greenbelt, MD 20771, USA}
\affiliation{Department of Astronomy, University of Maryland, College Park, MD 20742, USA}

\author[0000-0003-2302-9562]{Naoki Koshimoto}
\affiliation{Code 667, NASA Goddard Space Flight Center, Greenbelt, MD 20771, USA}
\affiliation{Department of Astronomy, University of Maryland, College Park, MD 20742, USA}

\author{Jean-Philippe Beaulieu}
\affiliation{School of Natural Sciences, University of Tasmania, Private Bag 37 Hobart, Tasmania, 70001, Australia}

\author[0000-0001-5860-1157]{Joshua W. Blackman}
\affiliation{School of Natural Sciences, University of Tasmania, Private Bag 37 Hobart, Tasmania, 70001, Australia}

\author[0000-0002-8131-8891]{Ian A. Bond}
\affiliation{School of Mathematical and Computational Sciences, Massey University, Auckland 0745, New Zealand}

\author[0000-0003-0303-3855]{Andrew A. Cole}
\affiliation{School of Natural Sciences, University of Tasmania, Private Bag 37 Hobart, Tasmania, 70001, Australia}

\author[0000-0001-9611-0009]{Jessica R. Lu}
\affiliation{Department of Astronomy, University of California Berkeley, Berkeley, CA 94701, USA}

\author[0000-0002-7901-7213]{Jean Baptiste Marquette}
\affiliation{Laboratoire d'Astrophysique de Bordeaux, Univ. Bordeaux, CNRS, B18N, all{\'e}e Geoffroy Saint-Hilaire, 33615 Pessac, France}
\affiliation{(Associated to) Sorbonne Universit{\'e}, UPMC et CNRS, UMR 7095, Institut d’Astrophysique de Paris, 98 bis Bd Arago, F-75014 Paris, France}

\author[0000-0003-2388-4534]{Cl\'ement Ranc}
\affiliation{Zentrum f{\"u}r Astronomie der Universit{\"a}t Heidelberg, Astronomiches Rechen-Institut, M{\"o}nchhofstr.\ 12-14, 69120 Heidelberg, Germany}

\author{Natalia Rektsini}
\affiliation{School of Natural Sciences, University of Tasmania, Private Bag 37 Hobart, Tasmania, 70001, Australia}

\author[0000-0002-9881-4760]{Aikaterini Vandorou}
\affiliation{Code 667, NASA Goddard Space Flight Center, Greenbelt, MD 20771, USA}
\affiliation{Department of Astronomy, University of Maryland, College Park, MD 20742, USA}

\correspondingauthor{S. K. Terry}
\email{sean.terry@berkeley.edu}

\begin{abstract}

\small \noindent We report new results for the gravitational microlensing target OGLE-2011-BLG-0950 from adaptive optics (AO) images using the Keck Observatory. The original analysis by \cite{choi:2012a} {and re-analysis by \cite{suzuki:2016a}}\, report degenerate solutions between planetary and stellar binary lens systems. This particular case is the most important type of degeneracy for exoplanet demographics, because the distinction between a planetary mass or stellar binary companion has direct consequences for microlensing exoplanet statistics. The 8 and 10-year baselines allow us to directly measure a relative proper motion of $4.20\pm 0.21\,$mas yr$^{-1}$, {confirming }the detection of the lens star system and {ruling }out the planetary companion models that predict a ${\sim}4 \times$ smaller relative proper motion. {The Keck data also rules out the wide stellar binary solution unless one of the components is a stellar remnant. } The combination of the lens brightness and close stellar binary light curve parameters yield primary and secondary star masses of $M_{\rm A} = 1.12^{+0.11}_{-0.09}M_{\sun}$ and $M_{\rm B} = 0.47^{+0.13}_{-0.10}M_{\sun}$ at a distance of $D_L = 6.70^{+0.55}_{-0.30}\,$kpc, and a projected separation of  $0.39^{+0.05}_{-0.04}\,$AU. {Assuming the predicted proper motions are measurably different}, the high resolution imaging method described here {can} be used to disentangle {this} degeneracy for events observed by the \textit{Roman} exoplanet microlensing survey using \textit{Roman} images taken near the beginning or end of the survey.
\\
\textit{Subject headings}: gravitational lensing: micro, binary systems \\
\end{abstract}


\section{Introduction} \label{sec:intro}
\indent Gravitational microlensing enables the detection of stars and exoplanets in a wide range of environments and distances along the line between Earth and the central region of the Galaxy. In addition to main sequence stars and exoplanets, more exotic systems like a Jupiter-analog orbiting a white dwarf \citep{blackman:2021a} and an isolated black hole or neutron star \citep{lam:2022a, sahu:2022a} have recently been published. \cite{koshimoto:2021b} recently used exoplanet microlensing detections to show there is no significant dependence on planet frequency with galactocentric distance. This implies planets residing in the Galactic Bulge are likely similar to planets near the Solar System.  \\
\indent While the number of detected microlensing events per year has been steadily rising, a unique circumstance of binary lens microlensing events is that they can possess different types of degeneracies. The well-known \textit{`close-wide'} degeneracy occurs when the central caustic shape between a closely separated binary lens and either of the two caustic shapes in a widely separated binary lens are essentially identical \citep{albrow:2001a}. There are degeneracies involving the binary mass ratio and finite size of the source star for low-magnification events, as well as degeneracies between planetary caustic perturbations and extreme flux ratio binary events \citep{gaudi:1998a}. A recent paper \citep{zhang:2022a} describes the \textit{‘offset’} degeneracy, which combines the close-wide degeneracy with other degeneracies relating models with different lens separations. Fortunately many of these degeneracies can be mitigated by obtaining high accuracy and well-sampled photometry during the microlensing light curve, particularly during caustic crossings or close approaches. Many of these degeneracies involve minor differences in the lens separation in the plane of the sky that are smaller than the uncertainty due to the unmeasured uncertainty along the line of sight, but in some cases, the degeneracy can involve very different separations. In the 30-event exoplanet demographics study of {\citet[hereafter S16]{suzuki:2016a}}, there were eight planetary systems with lens separation degeneracies that were {too} small to have an important effect on the demographics \citep{gould:2006a, bennett:2008a, bennett:2014a, bennett:2018a, janczak:2010a, bachelet:2012a, suzuki:2013a, nagakane:2017a}, two planetary events with strong close-wide degeneracies \citep{dong:2009a, fukui:2015a}, and the single event presented in this paper, where the {primary} degeneracy is between planetary and stellar binary models \citep{han:2008a}.  \\
\indent {\citet[hereafter C12]{choi:2012a}} {give further descriptions of} this degeneracy, which is related to the source star trajectory approaching a central caustic cusp due to either a planetary or stellar companion to the host. In the planetary case the source star with a large source radius crossing time ($t_*$) passes by the two strong cusps that bracket a negative perturbation region, at a trajectory angle $\alpha \sim 90\degree$ with respect to the planet-star axis. The effect of the weaker cusp between these two cusps is not obvious in the light curve. For the stellar binary case, the source star, with a much smaller $t_*$ passes two adjacent cusps that bracket a weaker negative perturbation region caused by a diamond- shaped caustic at a trajectory angle $\alpha \sim 45\degree$ with respect to the binary axis. This degeneracy is severe because regardless of how well the perturbation is sampled by the data, the interpretation of planetary or binary solution is generally limited by systematics of the photometry. In contrast to the degeneracies mentioned previously, this degeneracy is likely the most important of its kind when it comes to exoplanet demographics, because the ambiguity of the lens system parameters from the light curve modeling have a direct consequence on the exoplanet statistics that are drawn from the data (i.e. planetary systems vs. non-planetary systems). Finally, we refer to this degeneracy in a more specific manner than {C12}; we denote it the \textit{`central caustic cusp approach'} degeneracy. Since the source trajectory in both models approaches a caustic, finite source effects can be observed which allow the source radius crossing time, $t_*$, to be measured. With knowledge of $t_*$, an estimate of the lens-source relative proper motion, $\mu_{\textrm{rel}}$, can be made for each model. Further, high resolution follow-up observations can be made years after the event to directly measure $\mu_{\textrm{rel}}$ \citep{bennett:2015a, batista:2015a, bhattacharya:2018a, terry:2021a}, and compare the direct measurement to the light curve models to determine which interpretation is correct. {As we detail in Section \ref{sec:choi-compare}, C12 recognized the existence of systematic errors in the photometry from several datasets, which led them to conclude that a $\Delta\chi^2 {\sim} 105$ between the planetary and stellar binary solutions was not significant.} \\
\indent OGLE-2011-BLG-0950 is included in {S16} and \cite{suzuki:2018a}, which is one of the largest statistical studies of the microlensing exoplanet population. {S16 re-analyzed the event using optimized photometry from several datasets, which largely resolved the systematic photometry error problem that was present in the C12 analysis. This re-analysis resulted in a much larger uncertainty in the measurement of $t_*$ for the stellar binary models. The new modeling work of S16 and the current study show that the models with smaller and uncertain $t_*$ values are favored mainly by three data sets (with corresponding $\Delta \chi^2$); MOA ($\Delta \chi^2 \sim 57$), CTIO-I ($\Delta \chi^2 \sim 49$), and Danish ($\Delta \chi^2 \sim 63$). Further details about the differences between the C12, S16, and our new analysis are given in Section \ref{sec:prior-comparison}.} \\
\indent As mentioned previously, an ambiguous event like OGLE-2011-BLG-0950 is particularly important for populations statistics because ignoring or accepting a target like this could bias results in cases like {S16} that aimed to measure the cold exoplanet mass-ratio function. We note that another microlensing event (OGLE-2011-BLG-0526) exhibiting the same degeneracy was found in the same observing season as OGLE-2011-BLG-0950. {C12} claim this implies that this degeneracy may be common. Furthermore, a retrospective search through a nine-year sample (2006-2014)\footnote[9]{\href{https://exoplanetarchive.ipac.caltech.edu/docs/MOAMission.html}{\scriptsize https://exoplanetarchive.ipac.caltech.edu/docs/MOAMission.html}} of microlensing events from Microlensing Observations in Astrophysics (MOA; \citealt{bond01,sumi:2003a}) {yields at least three events (MOA-2012-BLG-201, MOA-bin-65, MOA-2014-BLG-051)} that show some evidence of this central caustic cusp approach degeneracy. We note that all of the events included in this 9-year sample are vetted and classified by-eye. Therefore, it may be the case that additional events exhibiting the central caustic cusp approach degeneracy are not identified in this 9-year sample. Lastly, it is expected that some fraction of the microlensing events that the upcoming \textit{Roman Galactic Exoplanet Survey} (RGES) discovers will exhibit this degeneracy. \\
\indent This paper is organized as follows: In Section \ref{sec:light-curve}, we perform improved photometry of the light curve {and compare our updated best-fit solutions with previous studies of the target}. In Section \ref{sec:follow-up} we describe the Keck adaptive optics (AO) follow-up analysis that confirms the stellar binary solution. Section \ref{sec:prop-motion} details our lens-source relative proper motion and flux ratio measurements from the 2019 and 2021 epochs. In Section \ref{sec:source_identification} and \ref{sec:lens_companion} we discuss the identification of the lens star and a subsequent search for a luminous lens star companion. We report the lens system physical parameters in Section \ref{sec:lens-properties}. Lastly, we discuss the overall results and conclude the paper in Section \ref{sec:conclusion}.


\section{Updated Light Curve Modeling} \label{sec:light-curve}
\begin{figure*}
\includegraphics[width=\linewidth]{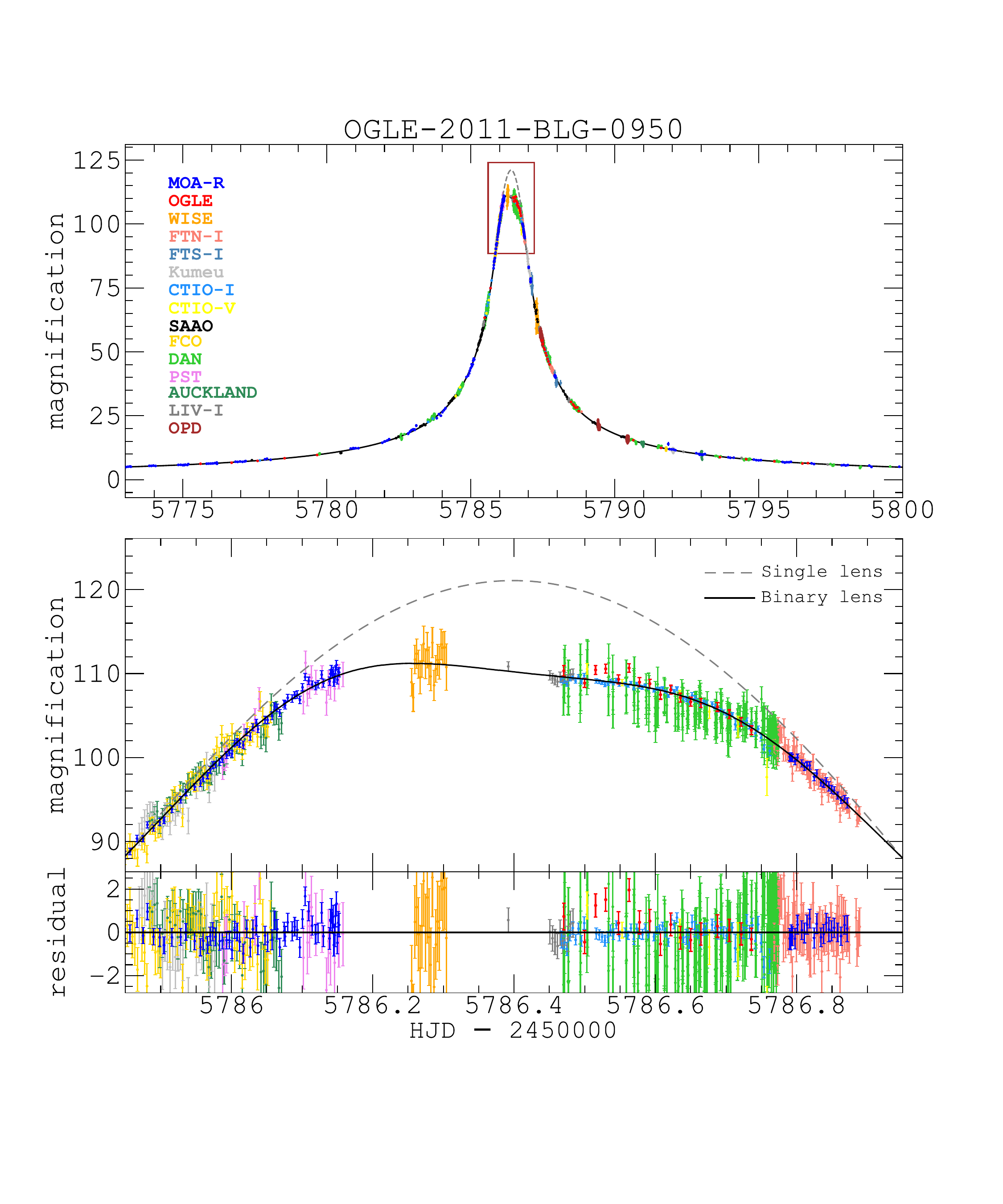}
\centering
\vspace{-3cm}
\caption{\footnotesize The OGLE-2011-BLG-0950 photometry with the updated best-fit light curve model (column two of table \ref{tab:lcparams}, the stellar binary with $s < 1$). {The middle panel shows an enlarged view of the peak, and the bottom panel shows the residuals to the best-fit stellar binary solution.} \label{fig:lc-photometry}}
\end{figure*}

The high magnification event OGLE-2011-BLG-0950/MOA-2011-BLG-336, located at RA $=$ 17:57:16.63, DEC $=$ -32:39:57.0 and Galactic coordinates ($l,b=(-1.93, -4.05)$) was alerted by the Optical Gravitational Lensing Experiment (OGLE; \citealt{udalski:1993a, udalski:2015a}) on 11th July 2011 and MOA on 31st July 2011. The perturbation was well sampled near the peak of the light curve, and a total of 15 telescopes performed observations at various times throughout the event. Telescopes in New Zealand (Auckland 0.4m, FCO 0.4m, Kumeu 0.4m), Chile (CTIO 1.3m, Danish 1.54m), Israel (WISE 1.0m), Australia (FTS 2.0m, {PST 0.3m}), and Hawaii (FTN 2.0m) performed follow-up observations around the high magnification peak on 13th August 2011. The measurements from each observatory can be seen in Figure \ref{fig:lc-photometry}, with the colored list of telescopes corresponding to the colors of each data point. \\
\indent There have been several improvements to the photometric reduction process since the {C12} analysis, therefore we have re-reduced the photometry for several data sets. We have used the updated photometry methods described in \citet{bond01,bond17} to reduce the data from the MOA 1.8m telescope, and the SMARTS telescope at CTIO. The SMARTS-CTIO data were reduced using difference imaging photometry \citep{bond01,bond17}, and the MOA data were corrected for errors due to chromatic differential refraction \citep{bennett:2012a, bond17}. The OGLE data have also been re-reduced and included in our new data sets. {This re-reduction procedure is similar to the re-analysis performed by \cite{suzuki:2016a}, which we describe further in Section \ref{sec:suzuki-compare}.}\\
\indent The updated light curve modeling follows the image-centered ray shooting method of \cite{bennett:1996a} and \cite{bennett:2010a}. The three fundamental microlensing parameters that are modeled for a single lens are the Einstein radius crossing time, $t_E$, the time and distance of closest approach between the source and lens center-of-mass, $t_0$ and $u_0$ respectively. For binary lenses, there are three additional parameters to model; the binary lens mass ratio, $q$, their separation, $s$, in units of the Einstein radius, and the angle between the source star trajectory and the binary lens axis, $\alpha$. As mentioned earlier, an additional parameter can be modeled if finite source effects are observed, this parameter is the source radius crossing time, $t_*$. The resulting best-fit models show the same four-fold degeneracy that {C12 and S16} describe. Our best-fit planetary and stellar binary solutions differ by $\Delta \chi^{2} \sim 27$, with nearly identical $\chi^2$ values for $s < 1$ and $s > 1$ within both solutions. {The degeneracy resulting from our updated modeling is more severe than what {C12} find ($\Delta \chi^{2} \sim 105$), and is in agreement with what {S16} find ($\Delta \chi^{2} \sim 20$). We discuss the differences between our results, the {C12} results, and the {S16} results in Section \ref{sec:prior-comparison}}. Figure \ref{fig:lc-photometry} shows the best-fit stellar binary model ($s < 1$) and Table \ref{tab:lcparams} shows the parameters of our best-fit close and wide models for both the planetary and stellar binary solutions. \\
\indent An estimate of the lens-source relative proper motion, $\mu_{\textrm{rel}}$ can be made if the angular size of the source can be determined. In order to measure the source radius, we need to determine the extinction corrected source magnitude and color. To achieve this, the SMARTS-CTIO $V$ and $I$ band data was calibrated to the OGLE-III catalog \citep{ogle3-phot}, and then we measured the red clump centroid at $V_{\rm rc} - I_{\rm rc} = 2.11$, $I_{\rm rc} =15.85$, following the method of \citet{bennett:2010b}. Using the bulge red clump giant magnitude, color, and distance from \citet{nataf:2013a}, we find $I$ and $V$ band extinction of $A_I = 1.33$ and $A_V = 2.16$. This gives an extinction corrected magnitude of $I_{s0} = 18.12 \pm 0.06$ and $V_{s0} = 19.00 \pm 0.07$. We then use a modified version of the surface brightness relations from
\citet{boyajian:2014a}, using stars spanning the range in colors that are relevant for microlensing targets:

\begin{equation}
    \textrm{log}({2\theta_{*}}) = 0.5014 + 0.4197(V_{s0} - I_{s0}) - 0.2 I_{s0}
\end{equation}

\begin{figure*}
\includegraphics[width=\linewidth]{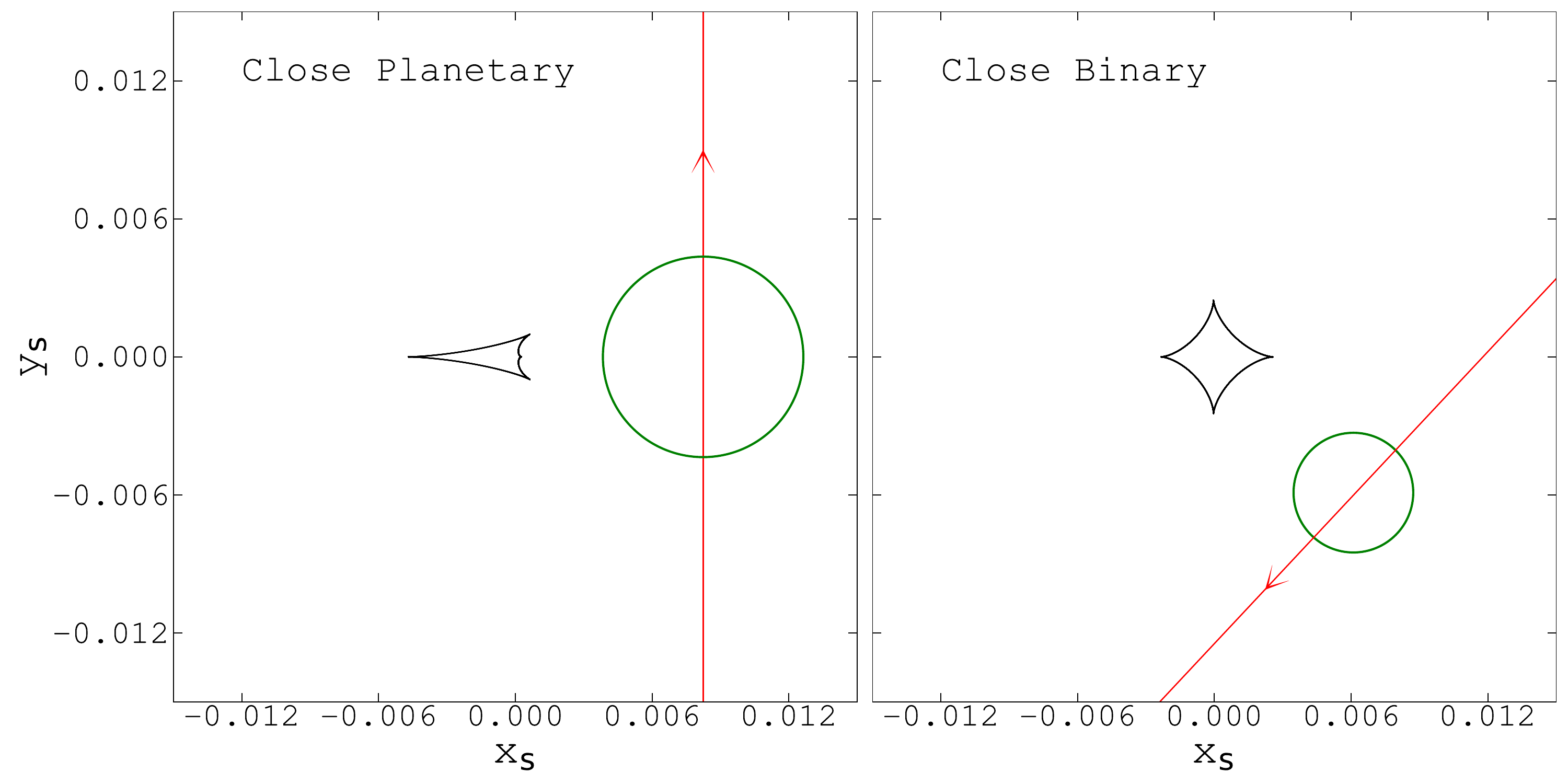}
\centering
\caption{\footnotesize Central caustic for the best planetary (left) and binary (right) models. The source size and trajectory are denoted by the green circle and red solid line, respectively. \label{fig:caustics}}
\end{figure*}

\noindent {This yields an angular source size of $\theta_{*} = 0.93 \pm 0.11\ \mu$as for the stellar binary solution, and $\theta_{*} = 0.90 \pm 0.10\ \mu$as for the planetary solution. Further, through the relation $\mu_{\textrm{rel}} = {\theta_{*}}/{t_{*}}$, we can determine the lens-source relative proper motion for both stellar binary and planetary interpretations. For the close stellar binary solution $\mu_{\textrm{rel,G}} = 3.95 \pm 4.10$ mas yr$^{-1}$, and for the wide planetary solution $\mu_{\textrm{rel,G}} = 1.05 \pm 0.20$ mas yr$^{-1}$, where the subscript G refers to the calculation being made in the inertial geocentric reference frame that moves with the Earth’s velocity at the time of the microlensing event. \\
\indent Figure \ref{fig:caustics} shows the central caustic for the close planetary and close binary models, along with the the source trajectory. The source size, $\rho_* = t_{*} / t_{E}$ (in $\theta_E$ units) is ${\sim} 2.7\times$ larger for the planetary model than for the stellar binary model because the magnification induced by the planetary model cusps is weaker than in the stellar binary case. Therefore the source must pass closer to the cusps to get the same signal. However, if the source passes closer to the cusps, this produces sharper light curve features, unless the $t_*$ value is increased to smooth them out. This is a generic feature of the degeneracy. Finally, our new light curve modeling results are consistent with the results of S16, and our results show smaller best-fit values for the mass ratios and larger $t_E$ values than what C12 report. These differences are carefully examined in Section \ref{sec:prior-comparison}. Details of the inferred physical parameters for the lens system are given in Section \ref{sec:lens-properties}.}

\begin{deluxetable*}{@{\extracolsep{5pt}}lcccc} [!htb]
\tablecaption{Best-fit Model Parameters \label{tab:lcparams}}
\setlength{\tabcolsep}{5.0pt}
\tablewidth{\columnwidth}
\tablehead
{
\colhead{} & \multicolumn{2}{c}{Stellar Binary} & \multicolumn{2}{c}{Planetary} \\
\cline{2-3} \cline{4-5}
\colhead{\hspace{-0.15cm}Parameter} & \colhead{s $<$ 1} & \colhead{s $>$ 1} & \colhead{s $<$ 1} & \colhead{s $>$ 1}
}
\startdata
$t_E$ (days) & $65.586 \pm 0.721$ & $101.870 \pm 8.195$ & $68.345 \pm 0.817$ & $68.153 \pm 0.752$ \\
$t_{0}$ (HJD$'$) & $5786.3965 \pm 0.0005$ & $5786.3925 \pm 0.0005$ & $5786.3969 \pm 0.0005$ & $5786.3959 \pm 0.0005$ \\
$u_0$ $(10^{-3})$ & $8.460 \pm 0.106$ & $5.443 \pm 0.431$ & $8.244 \pm 0.110$ & $8.612 \pm 0.101$ \\
$s$ & $0.0768 \pm 0017 $ & $21.9678 \pm 1.2473$ & $0.7257 \pm 0.0104$ & $1.3668 \pm 0.0191$ \\
$\alpha$ (rad) & $-0.719 \pm 0.005$ & $-0.718 \pm 0.004$ & $-1.531 \pm 0.003$ & $-1.532 \pm 0.002$ \\
$q$ & $0.417 \pm 0.115$ & $1.446 \pm 0.231$ & $(5.395 \pm 0.271) \times 10^{-4}$ & $(5.371 \pm 0.269) \times 10^{-4}$ \\
$t_*{}$ (days) & $0.0856 \pm 0.0882$ & $0.0635 \pm 0.0416$ & $0.3136 \pm 0.0041$ & $0.3129 \pm 0.0037$ \\
$I_S{}$ & $19.405 \pm 0.062$ & $19.397 \pm 0.061$ & $19.461 \pm 0.069$ & $19.457 \pm 0.067$ \\
$V_S{}$ & $21.144 \pm 0.072$ & $21.136 \pm 0.072$ & $21.199 \pm 0.077$ & $21.195 \pm 0.075$ \\
fit $\chi^2$ & $7046.21$ & $7047.68$ & $7020.62$ & $7019.52$
\enddata
\end{deluxetable*}

{
\subsection{Comparison to Previous Studies} \label{sec:prior-comparison}
\subsubsection{Choi et al. 2012 (C12)} \label{sec:choi-compare}
As mentioned previously, C12 performed the original light curve modeling for this event. This first modeling effort used many data sets {derived from earlier iterations of the photometric pipelines (OGLE, MOA, CTIO, and other $\mu$FUN datasets).} C12 was clearly aware of {systematics in some of those reductions}, as evidenced by their conclusion that the $\Delta\chi^2 {\sim}105$ favoring the planetary solution was not significant, when this large of a difference would typically be considered significant enough to rule out less likely solutions. The authors state that systematic residuals of the data from the planetary model are larger than the difference between the planetary and binary models. Our analysis using re-reduced photometry for several data sets has largely removed these systematic errors (following S16, Section \ref{sec:suzuki-compare}), which results in a significantly smaller $\Delta\chi^2 {\sim} 27$, only slightly favoring the planetary companion solutions. \\
\indent Further investigation of the C12 analysis shows evidence that several of the {earlier} photometric data sets contributed to their spurious measurement of $\rho_*$. This also led to {smaller} error bars on $\rho_*$ (see Table 1 of C12). We find that the finite source effect is largely unconstrained for the stellar binary models (see $t_*$ in Table \ref{tab:lcparams}) from our analysis of the re-reduced photometry. We also note that although C12 reports measurements of both components of the microlensing parallax, $\pi_{EE}$ and $\pi_{EN}$, the error bars for their estimates are of order ${\sim}100\%$, which we consider not a significant detection of $\pi_E$. We included parallax in our modeling and also do not find a significant measurement of $\pi_E$, the best-fit parallax values for the stellar binary case gives $\pi_{EE} = -0.040 \pm 0.034$ and $\pi_{EN} = -0.046 \pm 0.233$. \\
\indent Lastly, the best-fit $t_E$ that we find for the wide binary model, 101.9 days, is larger than the corresponding $t_E$ that C12 report. For the wide binary models, most of the light curve sees only the effect of one lens. So, the effective $t_E$ for the event is reduced by $\sqrt{1+q}$ or $\sqrt{1+\frac{1}{q}}$, depending on which star has a close approach with the source. While we use the same coordinate system for all of the models presented in Table \ref{tab:lcparams}, it appears C12 make a change in coordinate system for their wide binary model.
\subsubsection{Suzuki et al. 2016 (S16)} \label{sec:suzuki-compare} 
The light curve photometry was re-analyzed by S16 for their statistical study of the cold exoplanet population from MOA events detected between 2007 and 2012. Their analysis included the re-reduced data from the observatories listed in the previous section. Using this optimized photometry, S16 was able to remove many of the systematic photometry issues that were present in the C12 analysis, which resulted in many S16 results contradicting best-fit parameters reported by C12, particularly for the stellar binary solutions. S16 report a significantly smaller $\Delta\chi^2$ between the stellar binary and planetary solutions, a $\Delta\chi^2 {\sim}20$ that favors the planetary solutions. Our re-analysis of the light curve photometry follows that of the S16 analysis and gives $\Delta\chi^2 {\sim} 27$ between the stellar binary and planetary solutions. While this result is consistent with S16, we note that the target would have been formally classified as a planetary event in the S16 statistical analysis, given our $\Delta\chi^2 > 25$. However, an event very close to the S16 selection criteria warranted a careful investigation that included both possibilities in a Bayesian analysis (as S16 conducted).\\
\indent The use of optimized photometry led the S16 analysis to properly conclude a lack of constraint on $t_*$ in the stellar binary solutions. As can be seen in Table 3 of S16, the authors report uncertainties of ${\sim}100\%$ on the $t_*$ measurement for the stellar binary solutions. We find similarly large $t_*$ uncertainties in our best-fit values for the stellar binary solutions (Table \ref{tab:lcparams}). Finally, although the less certain $t_*$ values lead to larger errors on the $\mu_{\textrm{rel}}$ estimates for the binary models, the relative difference between these values and the planetary models $t_*$, as well as the inferred $\mu_{\textrm{rel}}$ is large enough (i.e. ${\sim}6\sigma$) such that the direct measurement of $\mu_{\textrm{rel}}$ with Keck (Section \ref{sec:prop-motion}) remains unambiguous (Figure \ref{fig:murel-posterior}).}

\section{Multi-Epoch High Resolution Imaging with Keck} \label{sec:follow-up}
The target OGLE-2011-BLG-0950 was observed with the NIRC2 instrument on Keck-II in the $K_\textrm{short}$ band ($\lambda_{c} = 2.146 \mu m$, hereafter Ks) on May 27, 2019. The target was also observed with the OSIRIS imager on Keck-I in the $K_\textrm{prime}$ band ($\lambda_{c} = 2.115 \mu m$, hereafter Kp) on July 14, 2021. The 2019 Ks band data have an average point spread function (PSF) full width at half maximum (FWHM) of 66.2 mas. The 2021 Kp band data have an average PSF FWHM of 66.8 mas, very similar to the 2019 Ks band data. Both epochs used the same tip/tilt guide star of R magnitude ${\sim}15$ at a separation of ${\sim}5.5 \arcsec$ to the north of the target. Although the 2019 NIRC2 data appear to be of equal or slightly better quality than 2021 OSIRIS, there are minor systematic artifacts on the PSF shape due to imperfect AO correcting on the NIRC2 system. These types of PSF systematics have been successfully modeled in the past on highly blended targets \citep{terry:2021a}. We regard the astrometry and photometry results from the 2019 data as reliable because they are consistent with the 2021 OSIRIS data (Section \ref{kpband-2021}). Ultimately, both datasets are consistent with the stellar binary interpretation for the lens system. The 2019 data suggested that we had detected the lens star at a proper motion only consistent with the stellar binary models, therefore we re-observed the target in 2021 to confirm that this star had the appropriate proper motion to be the lens system.
\begin{figure*}
\includegraphics[width=15cm]{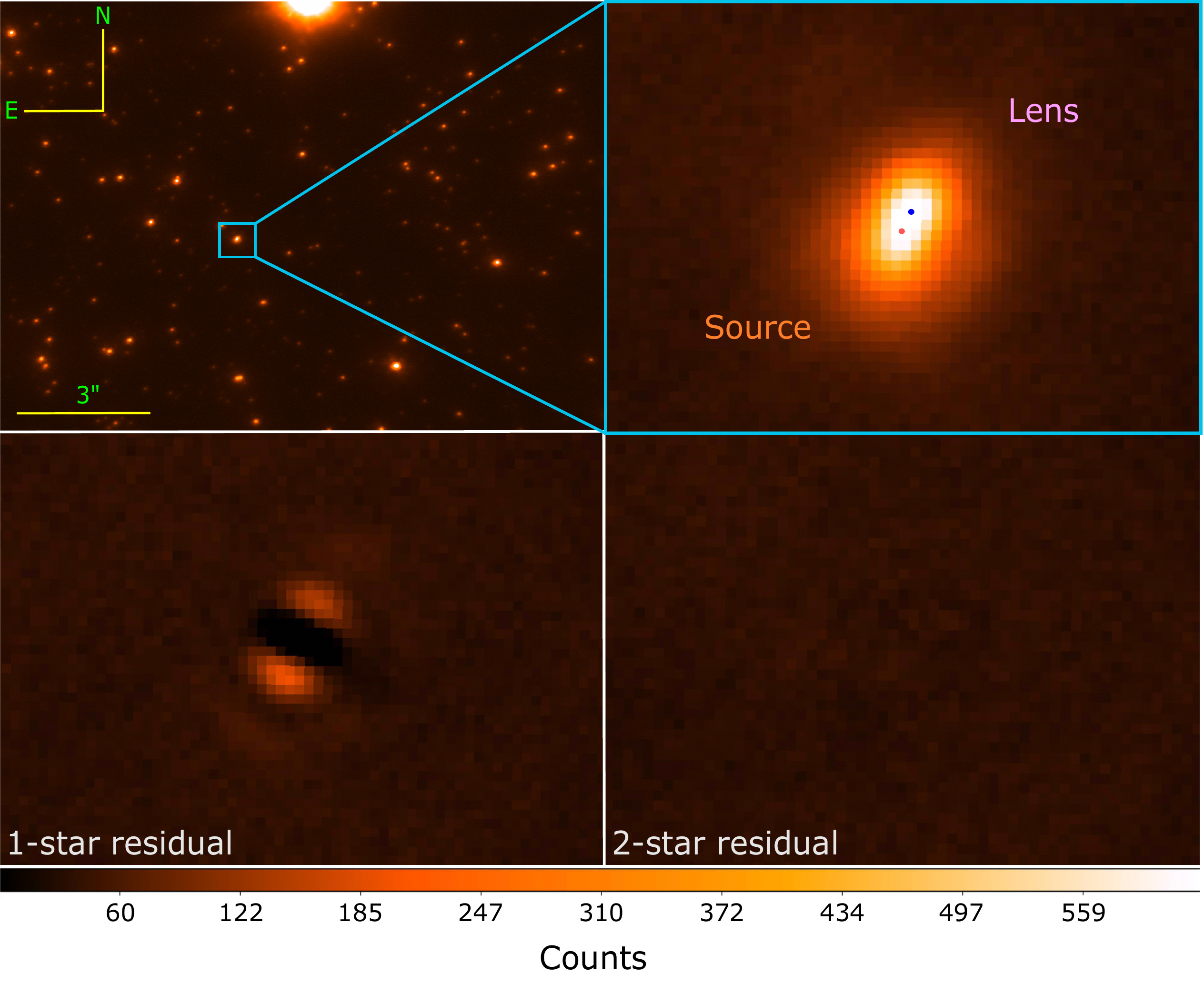}
\centering
\vspace{1mm}
\caption{\footnotesize \textit{Upper-left panel}: Co-added sum of 24 60-sec OSIRIS Kp band images from 2021. \textit{Upper-right panel}: closeup of blended source and lens stars. \textit{Lower-left panel}: The single-star PSF residual clearly showing a signal from the blend. \textit{Lower-right panel}: Two-star PSF residual showing a smooth subtraction. The color bar represents the intensity (counts) in the bottom panel residual images. \label{fig:quad-panel}}
\end{figure*}

\indent For the 2019 Ks band observations, both the NIRC2 wide and narrow cameras were used. The pixel scales for the wide and narrow cameras are 39.69 mas/pixel and 9.94 mas/pixel, respectively. Generally, the wide camera is used for photometric calibration to VVV (as described below), and the narrow camera is used to make the precise measurement of the lens and source star separations. All of the images were taken using the Keck-II laser guide star adaptive optics (LGSAO) system. For the narrow data, we combined 15 flat-field frames, six dark frames, and 15 sky frames for calibrating the science frames. A total of 21 Ks band science frames with an integration time of 60 seconds per frame were reduced using the Keck AO Imaging Data Reduction Pipeline (KAI) \citep{lu:2022a} to correct instrumental aberrations and geometric distortion \citep{ghez:2008a, lu:2008a, yelda:2010a, service:2016a}. Further, a co-add of 4 wide camera images were used for photometric calibration to images from the Vista Variables in the Via Lactea (VVV) survey \citep{minniti:2010a} following the procedures outlined in \cite{beaulieu:2018a}. The wide camera images were flat-fielded, dark current corrected, and stacked using the SWarp software \citep{bertin:2010a}. We performed astrometry and photometry on the co-added wide camera image using SExtractor \citep{bertin:1996a}, and subsequently calibrated the narrow camera images to the wide camera image by matching 80 bright isolated stars in the frames. The uncertainty resulting from this procedure is 0.07 magnitudes.\\
\indent For the 2021 Kp band data, we combined 40 flat-field frames, 10 dark frames, and 15 sky frames for calibrating our science images. The 24 Kp band science frames with an integration time of 60 seconds per frame were reduced with the same KAI pipeline. The combined frame can be seen on the upper-left panel of figure \ref{fig:quad-panel}, which has a PSF FWHM of $\sim$67 mas. It is worth noting the astrometric distortion solution for the OSIRIS imager has not yet been made publicly available. In development of the distortion solution \citep{freeman:inprep}, it has been shown that the absolute distortion at the pixels located closest to the centroids of the source and lens are $[dx, dy] = (-0.219, 0.097)$ pix for the source and $[dx, dy] = (-0.236, 0.102)$ pix for the lens (M. Freeman, private comm). The pixel scale for the OSIRIS imager is $9.95$ mas/pixel, so the difference in the measured distortion at these two locations translates to 0.17 mas on-sky. This is significantly smaller than the astrometric errors calculated from the \textit{DAOPHOT\_MCMC}+Jackknife analysis (Section \ref{sec:jackknife-mcmc}, Table \ref{table:epoch_seps}), thus we conclude the 2021 OSIRIS astrometry is not significantly affected by unaccounted for geometric distortions.

\subsection{PSF Fitting}\label{sec:photometry}
\begin{figure*}[!ht]
\includegraphics[width=\linewidth]{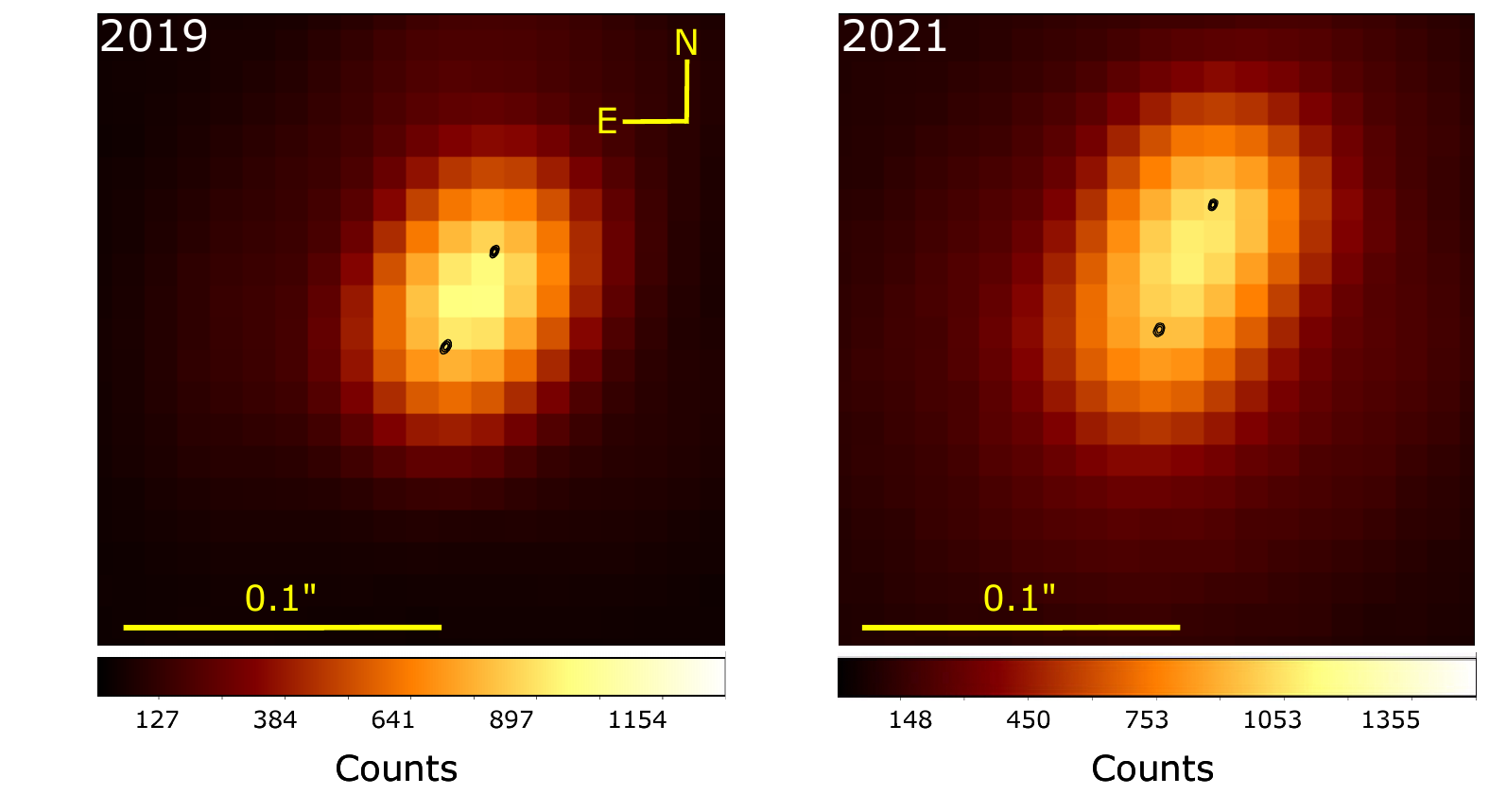}
\centering
\vspace*{3mm}
\caption{\footnotesize The best fit MCMC contours (68.3\%, 99.5\%, 99.7\%) for the source and lens positions are shown (in black) over-plotted on the 0.2"$\times$0.2" Ks band image from 2019 (left), and Kp band image from 2021 (right). The color bar refers to the pixel intensity. The measured separations and uncertainties in both epochs are given in Table \ref{table:epoch_seps}. This multi-epoch data confirms that the lens and source are separating from each other at a rate of $4.20 \pm 0.21$ mas yr$^{-1}$ in the Heliocentric reference frame. \label{fig:2019-2021}}
\end{figure*}

{In the binary solutions,} since the lens and source stars have a predicted separation of $0.49\times$FWHM and $0.65\times$FWHM in 2019 and 2021 respectively, we expect the stars to be partially resolved, so it is necessary to use a PSF fitting routine to measure both targets separately. Following the methods of \cite{bhattacharya:2018a} and \cite{terry:2021a}, we use a modified version of the DAOPHOT-II package \citep{stetson:1987a}, that we call \textit{DAOPHOT\_MCMC}, to run Markov Chain Monte Carlo (MCMC) sampling on the pixel grid encompassing the blended targets. Further details of the MCMC routine are given in \cite{terry:2021a}.
\\
\indent The stellar profile is clearly extended in both the NIRC2 and OSIRIS data, as can be seen in Figures \ref{fig:quad-panel} and \ref{fig:2019-2021}. Using \textit{DAOPHOT\_MCMC} to fit a single-star PSF to the blend produces the residual seen in the lower-left panel of Figure \ref{fig:quad-panel}, which shows a strong signal due to the extended flux from the blended star. Re-running the routine in the two-star fitting mode produces a significantly better fit as expected, with a $\chi^2$ improvement of $\Delta\chi^2 = 2462$. The two-star residual is nearly featureless, as can be seen in the lower-right panel of Figure \ref{fig:quad-panel}. Table \ref{table:dual-phot} shows the calibrated magnitudes for the two stars of {$K_{SSE}=17.02 \pm 0.05$ and $K_{NNW}=16.83 \pm 0.05$, where the subscript SSE represents the south-southeast star and NNW represents the north-northwest star.} The uncertainties are derived from the ``MCMC+Jackknife method" described in Section~\ref{sec:jackknife-mcmc}. Using the VVV extinction calculator \citep{gonzalez:2011a} and the \cite{nishiyama:2009a} extinction law, we find a K band extinction of $A_K = 0.20 \pm 0.06$. From our re-analysis of the light curve modeling (Section \ref{sec:light-curve}), we measure a source color of $(V - I)_S = 1.74 \pm 0.09$, which leads to an extinction-corrected color of $(V - I)_{S,0} = 0.88 \pm 0.09$. Further, we use the color-color relations of \cite{kenyon:1995a} and the I-band magnitude, $I_S = 19.405$ to predict a source K band magnitude of $K_S = 17.16 \pm 0.08$. This predicted magnitude is fainter than both stars detected in the Keck epochs, at ${\sim}1\sigma$ fainter than the SSE star and ${\sim}5\sigma$ fainter than the NNW star. While the predicted K band source magnitude is roughly consistent with the SSE star being the source, it is not as definitive as the typical result for this procedure because the lens and source usually have measurably different brightnesses \citep{bennett:2020a,bhattacharya:2021a,terry:2021a}. Because of this potential ambiguity, we perform an additional analysis using a Galactic model to confirm our {tentative} identification of the source star (Section \ref{sec:source_identification}).

\begin{deluxetable}{lcr}[!htb]
\deluxetablecaption{2021 PSF Photometry\label{table:dual-phot}}
\tablecolumns{3}
\setlength{\tabcolsep}{10.0pt}
\tablewidth{\linewidth}
\tablehead{
\colhead{\hspace{-15mm}Star} &
\colhead{Passband} & \colhead{\hspace{19mm}Magnitude}
}
\startdata
Lens & Keck $K$ & $16.83 \pm 0.07$\\
Source & Keck $K$ & $17.02 \pm 0.08$\\
Lens $+$ Source & Keck $K$ & $16.17 \pm 0.07$\\
\enddata
\tablenotetext{}{\footnotesize{\textbf{Note}. Magnitudes are calibrated to the VVV system, as described in section \ref{sec:follow-up}.}}
\end{deluxetable}

\subsubsection{Astrometric and Photometric Errors with MCMC+Jackknife}\label{sec:jackknife-mcmc}

\indent The full details of \textit{DAOPHOT\_MCMC} are given in \cite{terry:2021a}. Here we outline the new modifications we make that include the collation and propagation of astrometry/photometry errors through iterative MCMC runs on all jackknife frames. \\
\indent First, we generate the individual jackknife frames with the \texttt{reduce.jackknife()} function inside of the KAI pipeline. Then we run the standard \texttt{PSF} function inside \textit{DAOPHOT} on each jackknife frame to generate an empirical PSF associated with each jackknife frame. The same five reference stars within 4$\arcsec$ and 1 mag of the target were used in each jackknife frame to build the PSF model. It is necessary to generate a different empirical PSF for each jackknife frame because the shape of the PSF varies (sometimes significantly) between Keck AO images, and this PSF variation is precisely what we want to capture in the astrometric and photometric errors. We then employ an iterative scheme that runs \textit{DAOPHOT\_MCMC} on each jackknife frame. Finally, the output best-fit values and errors from each MCMC are combined in the Jackknife error calculation (i.e. equation 3 from \cite{bhattacharya:2021a}). These errors are reported for the astrometry and flux ratios in Tables \ref{table:epoch_seps} and \ref{tab:murel_fratio_compare}.

\subsection{2019 NIRC2 Analysis}\label{ksband-2019}
\indent As mentioned previously, the 2019 Ks band images have a PSF FWHM ${\sim}$ 66 mas, and the average Strehl Ratio (SR) across the 21 science frames is SR ${\sim}\ 0.33$. In an attempt to minimize the effect of PSF systematics, a careful selection of PSF reference stars were made to build the empirical PSF models for this epoch. In testing the different PSF models, we selected between four and nine reference stars with magnitudes $-1.0 < m < 1.0$ and separations $-5.5\arcsec < r < 5.5\arcsec$ from the target. In all cases, there remained significant correlated noise in the residuals after fitting and extracting sources with each candidate PSF model. The 2019 results reported in Section \ref{sec:prop-motion} and Table \ref{tab:murel_fratio_compare} come from an empirical PSF model built from five nearby stars, all of which are in common with PSF reference stars chosen for the 2021 OSIRIS PSF models, described in the next section.

\begin{deluxetable}{cccc}[!htb]
\deluxetablecaption{Measured Lens-Source Separations from 2019 and 2021\label{table:epoch_seps}}
\tablecolumns{3}
\setlength{\tabcolsep}{14.0pt}
\tablewidth{\linewidth}
\tablehead{
\colhead{}&
  \multicolumn{3}{c}{Separation (mas)} \\
\cline{2-4}
\colhead{Year} &
\colhead{East} & \colhead{North} & \colhead{Total}
}
\startdata
$2019$ & $14.11 \pm 0.40$ & $29.08 \pm 0.68$ & $32.32 \pm 0.79$\\
$2021$ & $16.65 \pm 0.27$ & $38.75 \pm 0.39$ & $42.18 \pm 0.47$\\
\enddata
\end{deluxetable}

\subsection{2021 OSIRIS Analysis}\label{kpband-2021}
The PSF FWHM for the 2021 OSIRIS data is comparable to the 2019 data, while the SR is measurably smaller for the 2021 data. This may be due to the difference in seeing for both nights, with an average seeing of 0.7$\arcsec$ for the 2019 epoch and 1.0$\arcsec$ for the 2021 epoch. Another possible reason for the $\Delta \textrm{SR} {\sim} 0.1$ might be the derivation of the SR itself on two independent AO systems (i.e. NIRC2 vs. OSIRIS). While a careful comparison of PSF metrics between the two imagers is compelling and probably worthwhile, it is beyond the scope of work in this paper. \\
\indent The best-fit separations, proper motions, and flux ratios for the 2021 epoch are given in Tables \ref{table:epoch_seps} and \ref{tab:murel_fratio_compare}. The 2021 best-fit results for both components of the lens-source relative proper motion are consistent with the 2019 results, and the best-fit flux ratio measured in 2021 differs from the 2019 result by ${\sim} 1\sigma$. The 2021 measurements confirm the lens identification from 2019, and both epochs confirm the stellar binary interpretation for the lens system (Section \ref{sec:prop-motion}). \\
\indent {Finally, we investigate the possibility that we have detected a luminous companion to the source star. We can infer the relative velocity between the two stars from the separation difference as measured in the 2019 and 2021 epochs (Table \ref{table:epoch_seps}). Using this information, the minimum source-companion velocity is calculated as: }

\begin{equation}
v_{sc} \,\,{\sim}\,\, \frac{10 \,\textrm{mas} \times 9 \,\textrm{kpc}}{2 \,\textrm{yr}} = 56 \,\textrm{AU/yr} \,\,{\sim}\,\, 8.9\times v_{\Earth}
\end{equation}

\noindent {Given the proportion $v \propto \frac{\sqrt{M + m}}{a}$ (Kepler's law), and assuming the source system total mass is $(M+m)\leq 3 M_{\sun}$, the semi-major axis for the source and its companion is $< 0.10$ AU. Considering that we measure a separation of 42 mas between the stars, this corresponds to a separation of ${\sim}$ 400 AU at 9 kpc, which rules out the scenario in which we are detecting a luminous companion to the source.}

\begin{deluxetable*}{lccc}[!htp]
\tablecaption{Best-Fit DAOPHOT-MCMC+Jackknife Results for Relative Proper Motion and Flux Ratio\label{tab:murel_fratio_compare}}
\tablecolumns{4}
\setlength{\tabcolsep}{20.5pt}
\tablewidth{\columnwidth}
\tablehead{
\colhead{Epoch} & \colhead{$\mu_{\textrm{rel,HE}}$ (mas yr$^{-1}$)} & \colhead{$\mu_{\textrm{rel,HN}}$ (mas yr$^{-1}$)} & \colhead{Flux Ratio (lens/source)}}
\startdata
2019 & $-1.811 \pm 0.204$ & $3.734 \pm 0.238$ & $1.13 \pm 0.09$\\
2021 & $-1.678 \pm 0.112$ & $3.906 \pm 0.240$ & $1.24 \pm 0.08$\\
Mean & $-1.745 \pm 0.117$ & $3.821 \pm 0.169$ & $1.19 \pm 0.06$\\
\enddata
\end{deluxetable*}


\section{Lens-Source Relative Proper Motion} \label{sec:prop-motion}
The 2019 and 2021 follow up observations were taken 7.79 and 9.92 years, respectively, after peak magnification in 2011. The motion of the source and lens on the sky is the primary cause for their apparent separation, however there is also a small component that can be attributed to the orbital motion of Earth. As this effect is of order $\leq0.10$ mas for a lens at a distance of $D_{L} \geq 6$ kpc, we are safe to ignore this contribution in our analysis as it is much smaller than the error bars on the stellar position measurements. The mean lens-source relative proper motion is measured to be $\mu_{\textrm{rel},H} = (\mu_{\textrm{rel,H,E}},\mu_{\textrm{rel,H,N}}) = (-1.745 \pm 0.117, 3.821 \pm 0.169)$ mas yr$^{-1}$, where `H' indications that these measurements were made in the Heliocentric reference frame, and the `E' and `N' subscripts represent the East and North on-sky directions respectively. Converting to Galactic coordinates, these proper motions are $\mu_{\textrm{rel,H,l}} = 1.016 \pm 0.117$ mas yr$^{-1}$ and $\mu_{\textrm{rel,H,b}} = 4.075 \pm 0.169$ mas yr$^{-1}$.

\begin{figure}
\includegraphics[width=\linewidth]{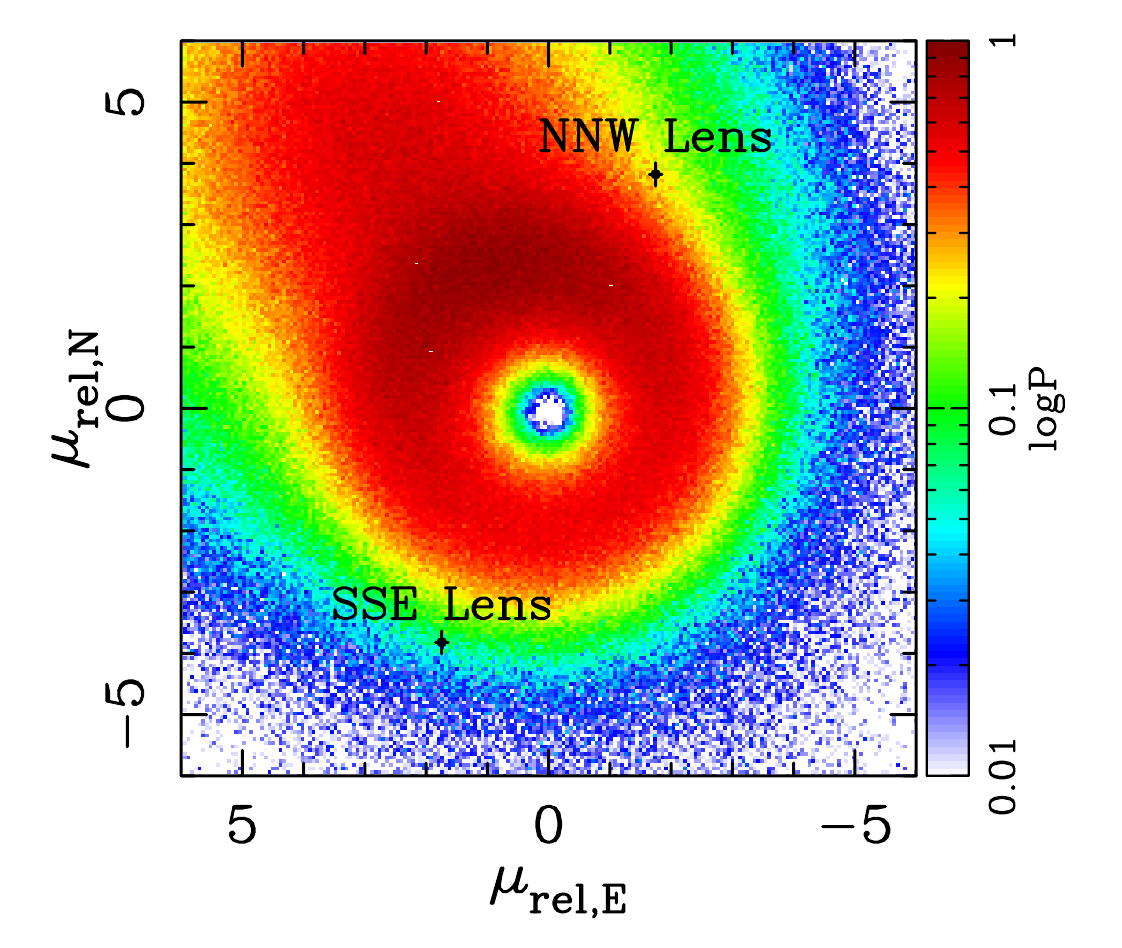}
\centering
\caption{\footnotesize Probability distribution for North and East components of lens-source relative proper motion ($\mu_{\textrm{rel}}$) using the Galactic model from \cite{koshimoto:2021a}. The lens positions (NNW and SSE) are plotted in black and are given by the relative motion of the two stars detected in the 2021 OSIRIS data. This implies the NNW star is $>3\times$ more likely to be the lens than the SSE star. \label{fig:mu_rel-2D-prob}}
\end{figure}

\indent Our light curve modeling (section \ref{sec:light-curve}) is performed in the Geocentric reference frame that moves with the Earth at the time of the event peak.
Thus, we must convert between the Geocentric and Heliocentric frames by using the relation given by \cite{dong:2009b}:

\begin{equation}\label{eq:mu-rel}
\mu_{\textrm{rel,H}} = \mu_{\textrm{rel,G}} + \frac{{\nu_{\Earth}}{\pi_{\textrm{rel}}}}{AU} \ ,
\end{equation}

\noindent where $\nu_{\Earth}$ is Earth's projected velocity relative to the Sun at the time of peak magnification. For OGLE-2011-BLG-0950 this value is $\nu_{\Earth \textrm{E,N}} = (12.223, -2.083)$ km/sec = $(2.574, -0.430)$ AU yr$^{-1}$ at HJD$' = 5786.40$. With this information and the relative parallax relation $\pi_{\rm{rel}} \equiv AU(1/D_{L} - 1/D_{S})$, we can express equation \ref{eq:mu-rel} in a more convenient form:

\begin{equation}
    \mu_{\textrm{rel,G}} = \mu_{\textrm{rel,H}} - (2.574, -0.430) \times (1/D_{L} - 1/D_{S})\, \textrm{mas yr$^{-1}$},
\end{equation}

\noindent where $D_{L}$ and $D_{S}$ are the lens and source distance, respectively, given in kpc. We have directly measured $\mu_{\textrm{rel,H}}$ from the Keck data, so this gives us the relative proper motion in the geocentric frame of $\mu_{\textrm{rel,G}} = 4.06 \pm 0.22\,$mas yr$^{-1}$. {While this proper motion is in agreement with the largely unconstrained stellar binary solution, $\mu_{\textrm{rel,G}} = 3.95 \pm 4.10\,$mas yr$^{-1}$, it strongly disagrees with the well-measured planetary solution, $\mu_{\textrm{rel,G}} = 1.05 \pm 0.20\,$mas yr$^{-1}$ from the light curve modeling}. Finally, the target identifications {and Keck-only separation measurements } that we have made between both epochs have confirmed the lens identification, as opposed to an unrelated non-lens star.

\begin{figure}
\includegraphics[width=\linewidth]{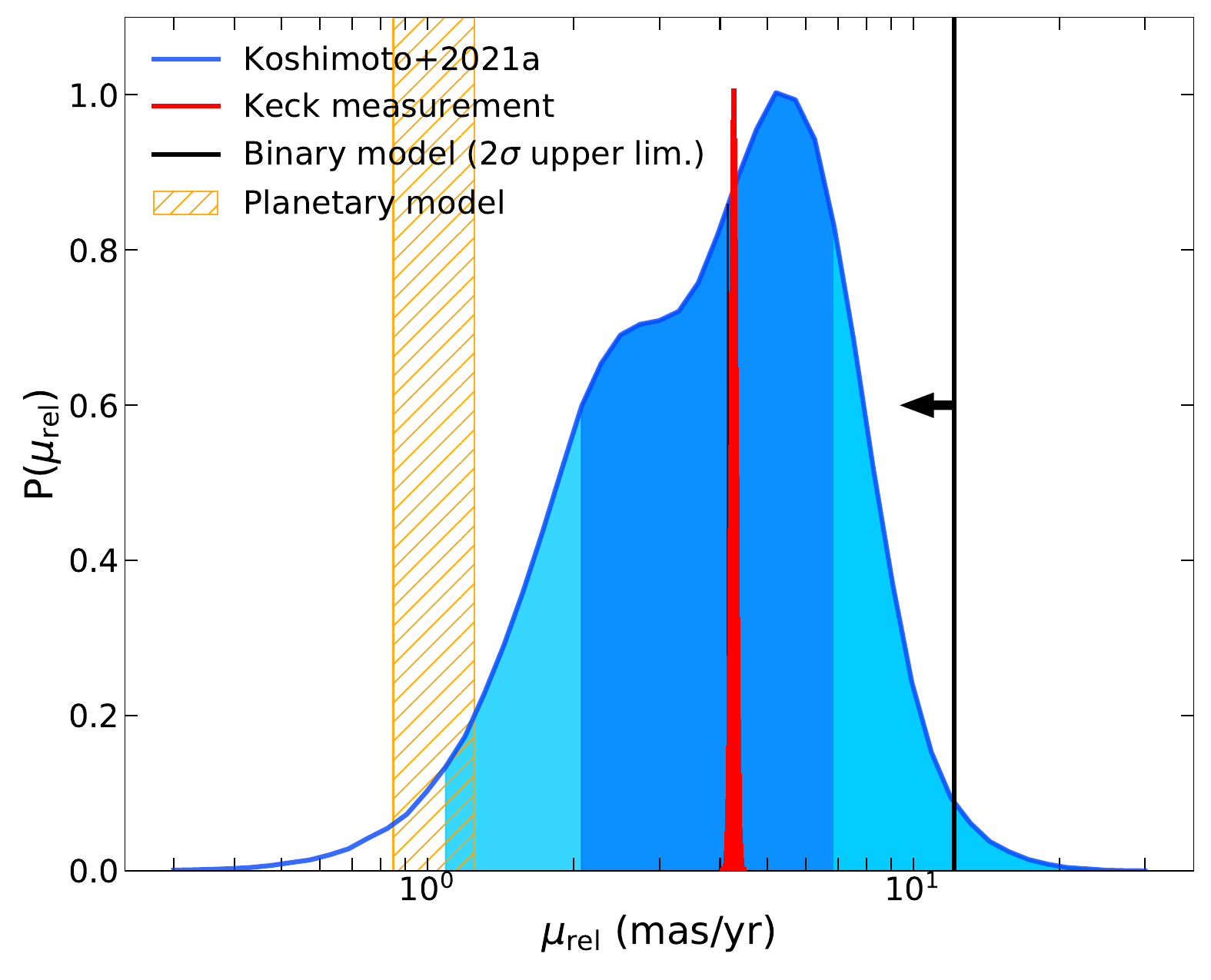}
\centering
\caption{\footnotesize Probability distribution for $\mu_{\textrm{rel}}$ derived using a Galactic model from \cite{koshimoto:2021a} (blue; dark shades are central 68\% of the distribution, light shades are central 95\% of the distribution). The posterior distribution from the direct measurement with Keck is shown in red. The orange hatchmark region indicates the predicted $\mu_{\textrm{rel}}$ from the best-fit planetary solution, {and black solid line with arrow shows the $2\sigma$ upper limit on $\mu_{\textrm{rel}}$ from the best-fit stellar binary solution. \label{fig:murel-posterior}}}
\end{figure}


\section{Source and Lens Star Identification}\label{sec:source_identification}
As described in Section \ref{sec:follow-up}, since the source and lens are similar brightness, the usual scheme to identify the source using color-color relations and the predicted K band source magnitude gives a less definitive identification in this case. Because of this, we calculated the 2D prior distribution of the lens-source relative proper motion using the \cite{koshimoto:2021a} Galactic model to determine which stars are the preferred source and lens. Figure \ref{fig:mu_rel-2D-prob} shows the results for this analysis, indicating the prior for each of the two possible lens stars. Assuming that the probability of having a companion star of a given mass ratio is constant and independent of the mass of the primary star and its position in our Galaxy, we calculated $\mu_{\rm rel}$ priors from the distribution of single lens stars that reproduces the Einstein radius crossing time that accounts for the primary star mass, i.e., $t_{\rm E}/\sqrt{1+q}$.\\
\indent This shows a preference for the NNW object in the Keck data to be the lens star(s) considering the stellar distribution in our Galaxy. The relative probability is $P_{NNW}/P_{SSE} = 3.79$ for the close binary scenario, this means the NNW object is $> 3\times$ more likely to be the lens than the SSE object. For the wide scenario, the NNW object is ${\sim}4\times$ more likely to be the lens than the SSE object. However, we note that for the wide solution the positions of the stars would be slightly different than the positions displayed in Figure \ref{fig:mu_rel-2D-prob}. The results of this analysis are consistent with what we find in Section \ref{sec:follow-up}; the predicted K band magnitudes imply the source star is the SSE object. Finally, it is indeed true that {we cannot completely rule out the possibility that the SSE star is the lens}. Nevertheless, our final results are little affected by this identification because both candidate stars are of similar brightness and thus the resulting lens properties will ultimately be similar. \\
\indent Figure \ref{fig:murel-posterior} shows the posterior probability distribution for the lens-source relative proper motion as directly measured by our Keck high-resolution data shown in red. Included in the figure is the prior probability distribution of relative proper motion derived using a Galactic model as described in \cite{koshimoto:2021a} and shown in blue. {The $\mu_{\textrm{rel,G}}$ estimate from the planetary model is given by the orange region, and the $2\sigma$ upper limit for the stellar binary model is given by the solid black line and arrow.} Although the best-fit planetary solution gives an unusually small $\mu_{\textrm{rel,G}} = 1.05 \pm 0.20$ mas yr$^{-1}$, it could not be completely ruled out until our direct measurement of $\mu_{\textrm{rel,G}} = 4.06 \pm 0.22$ mas yr$^{-1}$ with Keck confirmed the stellar binary solution.


\section{A Search for a Wide Lens Companion}\label{sec:lens_companion}
Since we have confirmed the stellar binary solution (Section \ref{sec:follow-up}) and identified the luminous lens star (Section \ref{sec:prop-motion}), we further investigate the possibility of resolving the companion to the primary lens. Given the best-fit close and wide solutions from the light curve modeling (Section \ref{sec:light-curve} and Table \ref{tab:lcparams}), the 2D projected separation between both lens components is ${\sim}0.1\times\theta_E$ for the close scenario and ${\sim}21.9\times\theta_E$ for the wide scenario. This translates to on-sky separations of 0.06 mas and 26.05 mas, respectively. Clearly the projected separation for the close stellar binary scenario is too small to be detected in either epoch, at a separation of ${\sim}0.001\times$FWHM. However, the projected separation for the wide binary scenario, of ${\sim}0.42\times$FWHM, should allow the companion to be detectable if it is luminous. We conduct two independent searches for this companion in the following sub-sections.

\subsection{Residual Pixel Grid}\label{sec:pixel-grid-resid}
We first analyze the residual pixel grid after the initial two-star PSF fitting is performed. The 2019 residual shows two over-subtracted regions and one under-subtracted region. The over-subtracted regions are both approximately $4{\times}4$ pixels in size and ${\sim}3.5\sigma$ below the mean pixel intensity inside the total PSF radius of 20 pixels. The under-subtracted region is approximately $5{\times}6$ pixels in size and $2.5\sigma$ above the mean pixel intensity inside the same PSF radius. The under-subtracted region may indicate the presence of an additional luminous source at this location, however its separation from the lens is inconsistent with the best wide binary solution from the light curve modeling. The location is measured at ${\sim} 78$ mas from the luminous lens component, and ${\sim} 69$ mas from the source, which are $3-4\times$ larger than the expected separation from the wide binary solution. As noted previously, the 2019 data has PSF systematics due to imperfect AO correcting. Similar over-subtracted and under-subtracted regions are observed in a majority of similar brightness stars in the central 5 arcseconds of the frame, particularly a similar under-subtracted region of similar size to that observed on the blended targets. There are no similarly blended stars of approximately equal magnitude nearby the target to perform a direct comparison of, so we focus this residual grid analysis on the target stars themselves. This analysis shows that the likely cause of the noise regions we detect in the 2019 residual is due to systematic errors on the PSF. \\
\indent In contrast, the 2021 residual pixel grid is significantly smoother, with at most ${\sim}0.8\sigma$ deviation above the mean pixel intensity within the same PSF radius as the 2019 data. Additionally, the empirical PSF models we derive for both the 2019 and 2021 Jackknife frames are uniformly circular. The largest non-uniformity in PSF shape used for any of the Jackknife frames is $(FWHM_Y - FWHM_X) = 0.023$ pix. For the close binary interpretation, two stars separated by ${\sim}0.001\times$FWHM are effectively located at the same pixel location in our data, which means the PSF fitting and extraction accounts for the combined flux of both lens stars. Lastly, we conclude that our search through the residual pixel grids show little to no definitive evidence for an additional source of flux at the expected distance from the luminous lens component for the wide binary scenario.

\subsection{Three-Star MCMC Search}\label{sec:3star-mcmc}
\begin{figure*}
  \includegraphics[width=1\textwidth]{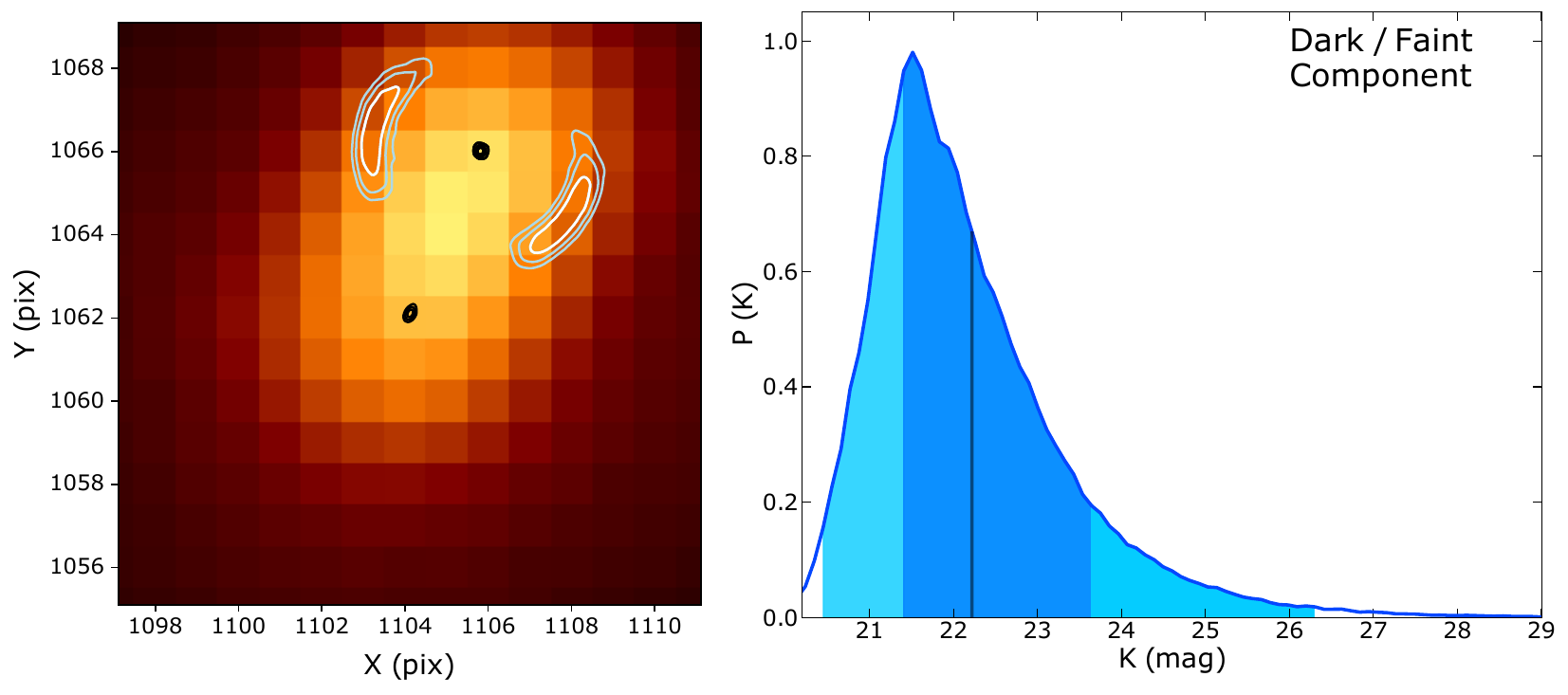}
\caption{\footnotesize \textit{Left}: Three-star MCMC distributions for the positions of the source star and luminous lens component (black contours) as well as the dark/faint lens component (blue/white contours). \textit{Right}: Posterior probability distribution for the K band magnitude of the dark/faint lens component. The central 68.3$\%$ of the distribution is shown in dark blue, and the remaining central 95.4$\%$ of the distribution is in light blue. \label{fig:3star-MCMC-magLim}}
\end{figure*}

We conduct a three-star \textit{DAOPHOT$\_$MCMC} search on the 2021 OSIRIS stacked frame with two constraints. The first constraint is the magnitude of the separation between star 2 (the luminous lens component), and star 3 (the non-luminous {or less-luminous} lens component). The second constraint we impose is the magnitude of the separation between star 1 (the source star), and star 3 (the {non/less-luminous} lens component). The latter constraint is imposed in order to prevent the MCMC from searching in locations that are disallowed by the wide binary best-fit parameters (Table \ref{tab:lcparams}). As mentioned in the previous section, there is evidence of PSF systematics in the 2019 NIRC2 data, therefore we chose to omit the 2019 epoch from the three-star MCMC analysis. For the 2021 OSIRIS data, we used a separation constraint of $2.65 \pm 0.25$ pix or $26.35 \pm 2.49$ mas for star 2/star 3 positions, and a separation constraint of $4.45 \pm 1.05$ pix or $44.24 \pm 10.44$ mas for the star 1/star 3 positions. \\
\indent The three-star MCMC results give positions of the source and luminous lens component that are in agreement (within 1$\sigma$) with our two-star MCMC analysis. The distribution for the position of the {non/less-luminous} lens component is shown in the left panel of Figure \ref{fig:3star-MCMC-magLim} as the white and blue contours. The possible positions for the undetected component are approximately perpendicular to the source-luminous lens component separation vector, as is required to be consistent with the best-fit wide binary solution from the light curve modeling. The right panel of Figure \ref{fig:3star-MCMC-magLim} shows the posterior distribution for the calibrated K band magnitude of the dark lens component, $K = 22.2^{+1.4}_{-0.8}$. This corresponds to a star 3/star 2 flux ratio of $f_{R} = 6.88^{+7.12}_{-4.92}\times10^{-3}$.\\
\indent {Lastly}, we compare the {\textit{DAOPHOT$\_$MCMC}} best-fit $\chi^2$ values between this three-star analysis and the two-star analysis (Section \ref{sec:photometry}). The two-star solution gives a marginally smaller best-fit $\chi^2$, with a difference of $\Delta \chi^2 {\sim}2$. Since the flux ratio between the lens components is very small in the three-star {\textit{DAOPHOT$\_$MCMC}} run, the result is a best-fit $\chi^2$ that is nearly identical to the two-star {\textit{DAOPHOT$\_$MCMC}} result. Additionally, we can compare the flux ratio distribution for the lens components {as given by the \textit{DAOPHOT$\_$MCMC} } analysis with the mass ratio that is {given by the best-fit wide binary solution from the light curve modeling (Section \ref{sec:light-curve})}. {Using empirical mass-luminosity relations \citep{henry:1993a,delfosse:2000a} with an assumed 0.1 magnitude uncertainty, we find nearly all flux ratios given by the \textit{DAOPHOT$\_$MCMC}} posterior distribution are inconsistent with the expected flux ratio of $f_{R}{\sim}0.44$ that corresponds to a mass ratio of $q{\sim}1.45$ from the wide binary solution (via light curve modeling). This allows us to rule out the wide binary scenario if we assume that both lens components are luminous.

\subsection{A Non-Luminous Primary Lens}\label{sec:dark-primary}
Although the analyses in Sections \ref{sec:pixel-grid-resid} and \ref{sec:3star-mcmc} suggest the close stellar binary scenario is preferred, we cannot completely rule out the possibility that a primary lens object or widely separated lens companion is below the detection threshold in the Keck data. If the {less massive component} is `dark' (i.e. a white dwarf (WD) or neutron star (NS)), then it would remain undetectable at any separation from its luminous counterpart. {For clarification, in the current Section and Section \ref{sec:dark-companion}, we use the term ``primary" to describe the (less-massive) lens component that the source trajectory comes nearest to (see $q > 1$ wide solution in Table \ref{tab:lcparams}).} \\
\indent Given the best-fit mass ratio from the light curve modeling, $q {\sim} 1.45$, and mass-luminosity relations \citep{henry:1993a} assuming all detected flux in Keck is from the companion star, then the companion mass is calculated as $M_B = 0.96^{+0.11}_{-0.09}M_{\sun}$. Further, this gives a primary lens mass of $M_A = 0.66^{+0.10}_{-0.08}M_{\sun}$. The lens system distance is calculated to be $D_L = 4.80^{+0.50}_{-0.30}\,$kpc for this case. This may be consistent with a white dwarf primary orbited by a main sequence companion.

\subsection{A Non-Luminous Companion}\label{sec:dark-companion}
\indent The alternative scenario to a dark primary lens is a dark companion orbiting a luminous primary lens. Again using the best-fit wide binary mass ratio and mass-luminosity relations assuming all of the measured flux is coming from the primary star, the primary lens mass is estimated to be $M_A = 1.08^{+0.11}_{-0.09}M_{\sun}$, orbited by a dark companion of mass $M_B = 1.57^{+0.15}_{-0.11}M_{\sun}$. For this scenario, the inferred lens system distance is {farther}, at $D_L = 5.79^{+0.61}_{-0.35}\,$kpc. The companion in this system may be consistent with a WD or NS, however it is very unlikely that a NS could form (through type II supernovae for example), and maintain a companion \citep{burrows:1987a}. We conduct a search of all public X-ray survey catalogues and find no objects in the vicinity of OGLE-2011-BLG-0950. The nearest unrelated transient object, XTE J1755-324, classified as an X-ray Nova \citep{revnivtsev:1998a}, is located 42{\arcmin } from the microlensing target. {We note that it is unlikely that a NS in a wide orbit would emit X-rays.} Further, the WD companion scenario would give an object that is at or above the Chandrasekhar limit. Such WDs are quite rare and do not remain stable for very long \citep{hillman:2016a}, further reducing the likelihood for the wide orbiting WD companion interpretation.\\
\indent To conclude, we find no strong evidence of an additional widely-separated lens object in either epoch. For the 2021 data, a three-star MCMC analysis gives a possible wide orbiting dark object with a brightness of $K = 22.2^{+1.4}_{-0.8}$, which is below the detection limit. Given the best-fit mass ratio for the wide binary model, this implies either a WD primary lens with a main sequence companion, or a main sequence primary lens with a WD/NS companion. Nevertheless, both wide stellar binary scenarios are less-preferred than the close stellar binary solution by the {\textit{DAOPHOT$\_$MCMC}} best-fit $\chi^2$, expected flux ratio, and WD/NS formation scenarios.


\section{Lens System Properties} \label{sec:lens-properties}

\begin{figure}
\includegraphics[width=\linewidth]{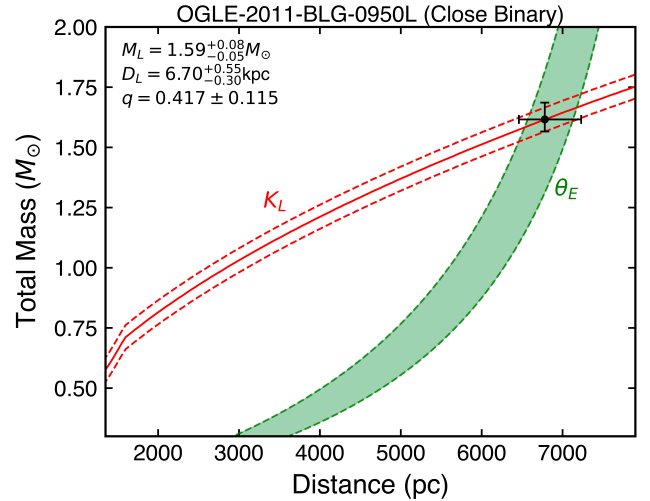}
\centering
\caption{\footnotesize The close stellar binary mass-distance relation for OGLE-2011-BLG-0950L with constraints from the {Keck K band lens flux measurement in red and the angular Einstein radius from the direct measurement of $\mu_{\textrm{rel}}$ in green. The mass-distance estimate given by the microlensing parallax from the light curve modeling is given in blue.} The individual masses are measured to be $M_{\rm A} = 1.12^{+0.11}_{-0.09}M_{\sun}$ and  $M_{\rm B} = 0.47^{+0.13}_{-0.10}M_{\sun}$. \label{fig:mass-dist}}
\end{figure}

As a result of the new direct measurement of the lens-source relative proper motion, we have successfully broken the central caustic cusp approach degeneracy for this event. The original four-fold degeneracy has now become a single degeneracy between the close and wide stellar binary solutions. Further, the results of our search for a luminous lens companion (Section \ref{sec:lens_companion}) give strong evidence that the close stellar binary interpretation is the correct solution. Working from this point, we use the Keck lens flux and mass-luminosity relations \citep{henry:1993a,henry:1999a,delfosse:2000a} in order to constrain the stellar binary lens distance. Given a Galactic latitude of $b = -4.05\degree$ and lens system distance of ${\sim}7$kpc, the lens is likely to be behind most of the interstellar dust that is in the foreground of the source. We can describe the foreground extinction as follows:

\begin{equation}
    A_{i,L} = \frac{1 - e^{-|D_{L}(\textrm{sin}b)/h_{\textrm{dust}}|}}{1 - e^{-|D_{S}(\textrm{sin}b)/h_{\textrm{dust}}|}}A_{i,S}
\end{equation}

\noindent where $i$ represents the passbands; $I, V$, and $K$. We assume a dust scale height of $h_{\textrm{dust}} = 0.10\pm 0.02\,$kpc. Additionally, we can use the $\theta_E$ value {inferred from the direct measurement of $\mu_{\textrm{rel}}$}, along with a mass-distance relation assuming we know the distance to the source \citep{bennett:2008a, gaudi:2012a}:

\begin{equation}
    M_{L} = \frac{c^2}{4G}\theta_{E}^{2}\frac{D_{S}D_{L}}{D_{S}-D_{L}},
\end{equation}

\noindent where $M_{L}$ is the lens mass, $G$ and $c$ are the gravitational constant and speed of light. $D_{L}$ and $D_{S}$ are the distance to the lens and source, respectively. Figure \ref{fig:mass-dist} shows the measured mass and distance of the binary lens. The red curve represents the constraint from the mass-luminosity relation, with dashed lines representing the error from the Keck lens flux measurement. For this close stellar binary case, the empirical mass-luminosity relation was numerically calculated for each star in the lens system, considering that the measured lens flux with Keck is the combination of two luminous stars with a non-negligible mass ratio $q$. Additionally the $\theta_{E}$ constraint {from the direct measurement of $\mu_{\textrm{rel}}$ from Keck is shown in green, with dashed lines representing the error on the $\theta_{E}$ measurement. Lastly, we include in Figure \ref{fig:mass-dist} the estimated mass and distance values from the largely unconstrained microlensing parallax measurement as given by the stellar binary light curve modeling.} \\
\indent The OGLE-2011-BLG-0950 lens system is located at a distance of ${\sim}6.7$ kpc and has a mass ratio of $q = 0.42 \pm 0.12$ and total mass of $M_{TOT} = 1.59^{+0.08}_{-0.05}M_{\sun}$. {The combination of the relatively large error on the mass ratio from the light curve modeling and the relatively small error on the total mass from the Keck imaging gives a primary mass of $M_A = 1.12^{+0.11}_{-0.09}M_{\sun}$ and secondary mass of $M_B = 0.47^{+0.13}_{-0.10}M_{\sun}$}. These masses are consistent with a $K$ dwarf orbiting a star near the top of the main sequence. The 2D projected separation for the close binary is measured to be $a_{\perp} = 0.39^{+0.05}_{-0.04}\,$AU for the binary stars. Table \ref{tab:lens-params} gives all of the lens system parameters along with their 2$\sigma$ ranges for the close stellar binary solution.\\

\begin{deluxetable*}{lccc}[!htp]
\deluxetablecaption{Close Stellar Binary Lens System Properties \label{tab:lens-params}}
\tablecolumns{4}
\setlength{\tabcolsep}{14.0pt}
\tablewidth{\columnwidth}
\tablehead{
\colhead{\hspace{-6cm}Parameter} & \colhead{Units} &
\colhead{Values \& RMS} & \colhead{2-$\sigma$ range}
}
\startdata
Angular Einstein Radius ($\theta_E$) & mas & $0.76 \pm 0.08$ & $0.61-0.92$\\
Geocentric lens-source relative proper motion ($\mu_{\textrm{rel,G}}$) & mas yr$^{-1}$ & $4.06 \pm 0.22$ & $3.62-4.50$\\
Primary Mass ($M_{\rm A}$) & $M_{\Sun}$ & $1.12^{+0.11}_{-0.09}$ & $0.94-1.34$\\
Secondary Mass ($M_{\rm B}$) & $M_{\Sun}$ & $0.47^{+0.13}_{-0.10}$ & $0.28-0.74$\\
2D Separation ($a_{\perp}$) & AU & $0.39^{+0.05}_{-0.04}$ & $0.31-0.49$\\
Lens Distance (D$_{L}$) & kpc & $6.70^{+0.55}_{-0.30}$ & $6.10-7.80$\\
Source Distance (D$_{S}$) & kpc & $9.17^{+1.07}_{-0.45}$ & $8.27-11.31$\\
\enddata
\end{deluxetable*}


\section{Discussion and Conclusions} \label{sec:conclusion}

\indent Our follow-up high resolution observations of the microlensing target OGLE-2011-BLG-0950 have allowed us to make a direct measurement of flux from the lens star(s) as well as a precise determination of the direction and amplitude of the lens-source relative proper motion. We are able to successfully break the central caustic cusp approach degeneracy by showing the lens-source relative proper motion directly measured with Keck is only compatible with the stellar binary solutions for the lens system. Further, the probability distribution estimates for the lens-source relative proper motion derived using a Galactic model show very low $\mu_{\textrm{rel}}$ values for the planetary companion models. This low probability for the planetary $\mu_{\textrm{rel}}$ values could have been noticed a priori. Ultimately, the subsequent Keck observations have now completely ruled out any planetary companion models. Additionally, an analysis of the 2021 OSIRIS data and its residuals favor the close stellar binary solution, however we cannot fully rule out the wide binary scenario if the binary includes a stellar remnant. \\
\indent We modified the PSF fitting routine \textit{DAOPHOT\_MCMC} \citep{terry:2021a} to calculate Jackknife (i.e. drop-one frame) errors for astrometry and photometry. We used these \textit{DAOPHOT\_MCMC}+Jackknife error bars for our final analysis. These pipelines, or something similar, can be used in future analyses of highly blended microlensing targets, and will likely form the foundation for the \textit{Roman} mass measurement method. \textit{Roman} will collect the precursor or follow-up data that is needed to enable direct lens detections. The maximum time baseline between two \textit{Roman} epochs will be ${\sim} 5$ years. We note that \cite{bennett:2006a} and \cite{dong:2009a} measured the lens-source separations (with \textit{HST}) for the first two planetary microlensing events less than two years after the events, these measurements were recently confirmed by \cite{bennett:2020a} and \cite{bhattacharya:inprep}. With at least $100\times$ more images during its microlensing survey, \textit{Roman} should reliably measure lens-source separations for many detected events.  \\
\indent {The study of {S16} includes a re-analysis of OB110950 using optimized photometry, in which they report the same four-fold degeneracy found in C12. Importantly, the S16 modeling find the correct best-fit parameters for the stellar binary models, which we confirm in this work. To avoid biasing the statistical results, S16 include both stellar binary and planetary companion possibilities in a Bayesian analysis, they formally conclude that the planetary companion solution is slightly ($\Delta\chi^2 {\sim} 20$) preferred over the stellar binary solution. Our new results that firmly place OB110950 in the stellar binary regime would at most have only a minor {effect} (${<}\,4$\%) on the broken power-law function that was reported in {S16}.}\\
\indent Finally, as described in {C12}, this degeneracy is severe in the sense that two significantly different-shaped central caustics can lead to indistinguishable light curve perturbations {(C12 report their $\Delta\chi^2 {\sim}105$ is not significant because of systematics in the photometry). The re-analysis of the light curve by S16 and this work shows that the degeneracy becomes more severe with the use of newly re-reduced photometry.} The fact that {C12} identified two events from a single observing season with this degeneracy, in addition to {${\sim}3$ known events} in a nine-year MOA sample that also show evidence of this degeneracy, implies that this type of degeneracy may be common and will likely be encountered in a non-negligible fraction of \textit{Roman} microlensing detections. The techniques described in this work should allow for {many} of the events exhibiting this degeneracy to be reconciled, {given measurably different $\mu_{\textrm{rel}}$ values for the degenerate solutions.}

The Keck Telescope observations and data analysis were supported by a NASA Keck PI Data Award, 80NSSC18K0793, administered by the NASA Exoplanet Science Institute. Data presented herein were obtained at the W. M. Keck Observatory from telescope time allocated to the National Aeronautics and Space Administration through the agency's scientific partnership with the California Institute of Technology and the University of California. The Observatory was made possible by the generous financial support of the W. M. Keck Foundation. The authors wish to recognize and acknowledge the very significant cultural role and reverence that the summit of Maunakea has always had within the indigenous Hawaiian community. We are most fortunate to have the opportunity to conduct observations from this mountain. DPB and AB were supported by NASA through grant NASA-80NSSC18K0274. This work was supported by the University of Tasmania through the UTAS Foundation and the endowed Warren Chair in Astronomy and the ANR COLD-WORLDS (ANR-18-CE31-0002). NK was supported by the JSPS overseas research fellowship. Work by CR was supported by a Research fellowship of the Alexander von Humboldt Foundation. IAB is supported by a Marsden Grant from the Royal Society of New Zealand (MAU-1901). This research was also supported in part by the Australian Government through the Australian Research Council Discovery Program (project number 200101909) grant awarded to Cole and Beaulieu. Some of this research has made use of the NASA Exoplanet Archive, which is operated by the California Institute of Technology, under the Exoplanet Exploration Program. This work made use of data from the Astro Data Lab at NSF's OIR Lab, which is operated by the Association of Universities for Research in Astronomy (AURA), Inc. under a cooperative agreement with the National Science Foundation. The authors thank Dr. Matthew Freeman for providing OSIRIS distortion details. SKT thanks Gary Bennett for helpful software discussions and comments on this manuscript. {The authors would like to thank the anonymous referee for helpful comments that improved the structure and led to a stronger manuscript.}

\textit{Software}: daophot\_mcmc \citep{terry:2021a}, genulens \citep{koshimoto:code}, KAI \citep{lu:2022a}, Matplotlib \citep{hunter:2007a}, Numpy \citep{oliphant:2006a}.

\bibliographystyle{aasjournal}
\bibliography{Terry_ob110950.bib}

\begin{thebibliography}{}
\expandafter\ifx\csname natexlab\endcsname\relax\def\natexlab#1{#1}\fi
\providecommand{\url}[1]{\href{#1}{#1}}

\bibitem[{Albrow {et~al.}(2001)Albrow, An, Beaulieu, Caldwell, DePoy, Dominik,
  Gaudi, Gould, Greenhill, Hill, {et~al.}}]{albrow:2001a}
Albrow, M.~D., An, J., Beaulieu, J.-P., {et~al.} 2001, ApJ, 549, 759

\bibitem[{Bachelet {et~al.}(2012)Bachelet, Shin, Han, Fouqu{\'e}, Gould,
  Menzies, Beaulieu, Bennett, Bond, Dong, {et~al.}}]{bachelet:2012a}
Bachelet, E., Shin, I.-G., Han, C., {et~al.} 2012, The Astrophysical Journal,
  754, 73

\bibitem[{{Batista} {et~al.}(2015){Batista}, {Beaulieu}, {Bennett}, {Gould},
  {Marquette}, {Fukui}, \& {Bhattacharya}}]{batista:2015a}
{Batista}, V., {Beaulieu}, J.~P., {Bennett}, D.~P., {et~al.} 2015, \apj, 808,
  170

\bibitem[{{Beaulieu} {et~al.}(2018){Beaulieu}, {Batista}, {Bennett},
  {Marquette}, {Blackman}, {Cole}, {Coutures}, {Danielski}, {Dominis Prester},
  {Donatowicz}, {Fukui}, {Koshimoto}, {Lon{\v{c}}ari{\'c}}, {Morales}, {Sumi},
  {Suzuki}, {Henderson}, {Shvartzvald}, \& {Beichman}}]{beaulieu:2018a}
{Beaulieu}, J.~P., {Batista}, V., {Bennett}, D.~P., {et~al.} 2018, \aj, 155, 78

\bibitem[{Bennett {et~al.}(2012)Bennett, Sumi, Bond, Kamiya, Abe, Botzler,
  Fukui, Furusawa, Itow, Korpela, {et~al.}}]{bennett:2012a}
Bennett, D., Sumi, T., Bond, I., {et~al.} 2012, The Astrophysical Journal, 757,
  119

\bibitem[{Bennett(2008)}]{bennett:2008a}
Bennett, D.~P. 2008, in Exoplanets (Springer), 47--88

\bibitem[{Bennett(2010)}]{bennett:2010a}
---. 2010, ApJ, 716, 1408

\bibitem[{{Bennett} {et~al.}(2006){Bennett}, {Anderson}, {Bond}, {Udalski}, \&
  {Gould}}]{bennett:2006a}
{Bennett}, D.~P., {Anderson}, J., {Bond}, I.~A., {Udalski}, A., \& {Gould}, A.
  2006, \apjl, 647, L171

\bibitem[{{Bennett} \& {Rhie}(1996)}]{bennett:1996a}
{Bennett}, D.~P., \& {Rhie}, S.~H. 1996, \apj, 472, 660

\bibitem[{{Bennett} {et~al.}(2010){Bennett}, {Rhie}, {Nikolaev}, {Gaudi},
  {Udalski}, {Gould}, {Christie}, {Maoz}, {Dong}, {McCormick}, {Szyma{\'n}ski},
  {Tristram}, {Macintosh}, {Cook}, {Kubiak}, {Pietrzy{\'n}ski},
  {Soszy{\'n}ski}, {Szewczyk}, {Ulaczyk}, {Wyrzykowski}, {OGLE Collaboration},
  {DePoy}, {Han}, {Kaspi}, {Lee}, {Mallia}, {Natusch}, {Park}, {Pogge},
  {Polishook}, {{\ensuremath{\mu}}FUN Collaboration}, {Abe}, {Bond}, {Botzler},
  {Fukui}, {Hearnshaw}, {Itow}, {Kamiya}, {Korpela}, {Kilmartin}, {Lin},
  {Ling}, {Masuda}, {Matsubara}, {Motomura}, {Muraki}, {Nakamura}, {Okumura},
  {Ohnishi}, {Perrott}, {Rattenbury}, {Sako}, {Saito}, {Sato}, {Skuljan},
  {Sullivan}, {Sumi}, {Sweatman}, {Yock}, {MOA Collaboration}, {Albrow},
  {Allan}, {Beaulieu}, {Bramich}, {Burgdorf}, {Coutures}, {Dominik}, {Dieters},
  {Fouqu{\'e}}, {Greenhill}, {Horne}, {Snodgrass}, {Steele}, {Tsapras},
  {PLANET}, {RoboNet Collaborations}, {Chaboyer}, {Crocker}, \&
  {Frank}}]{bennett:2010b}
{Bennett}, D.~P., {Rhie}, S.~H., {Nikolaev}, S., {et~al.} 2010, \apj, 713, 837

\bibitem[{{Bennett} {et~al.}(2014){Bennett}, {Batista}, {Bond}, {Bennett},
  {Suzuki}, {Beaulieu}, {Udalski}, {Donatowicz}, {Bozza}, {Abe}, {Botzler},
  {Freeman}, {Fukunaga}, {Fukui}, {Itow}, {Koshimoto}, {Ling}, {Masuda},
  {Matsubara}, {Muraki}, {Namba}, {Ohnishi}, {Rattenbury}, {Saito}, {Sullivan},
  {Sumi}, {Sweatman}, {Tristram}, {Tsurumi}, {Wada}, {Yock}, {MOA
  Collaboration}, {Albrow}, {Bachelet}, {Brillant}, {Caldwell}, {Cassan},
  {Cole}, {Corrales}, {Coutures}, {Dieters}, {Dominis Prester}, {Fouqu{\'e}},
  {Greenhill}, {Horne}, {Koo}, {Kubas}, {Marquette}, {Martin}, {Menzies},
  {Sahu}, {Wambsganss}, {Williams}, {Zub}, {PLANET Collaboration}, {Choi},
  {DePoy}, {Dong}, {Gaudi}, {Gould}, {Han}, {Henderson}, {McGregor}, {Lee},
  {Pogge}, {Shin}, {Yee}, {{\ensuremath{\mu}}FUN Collaboration},
  {Szyma{\'n}ski}, {Skowron}, {Poleski}, {Koz{\l}owski}, {Wyrzykowski},
  {Kubiak}, {Pietrukowicz}, {Pietrzy{\'n}ski}, {Soszy{\'n}ski}, {Ulaczyk},
  {OGLE Collaboration}, {Tsapras}, {Street}, {Dominik}, {Bramich}, {Browne},
  {Hundertmark}, {Kains}, {Snodgrass}, {Steele}, {RoboNet Collaboration},
  {Dekany}, {Gonzalez}, {Heyrovsk{\'y}}, {Kandori}, {Kerins}, {Lucas},
  {Minniti}, {Nagayama}, {Rejkuba}, {Robin}, \& {Saito}}]{bennett:2014a}
{Bennett}, D.~P., {Batista}, V., {Bond}, I.~A., {et~al.} 2014, \apj, 785, 155

\bibitem[{{Bennett} {et~al.}(2015){Bennett}, {Bhattacharya}, {Anderson},
  {Bond}, {Anderson}, {Barry}, {Batista}, {Beaulieu}, {DePoy}, {Dong}, {Gaudi},
  {Gilbert}, {Gould}, {Pfeifle}, {Pogge}, {Suzuki}, {Terry}, \&
  {Udalski}}]{bennett:2015a}
{Bennett}, D.~P., {Bhattacharya}, A., {Anderson}, J., {et~al.} 2015, \apj, 808,
  169

\bibitem[{Bennett {et~al.}(2018)Bennett, Udalski, Bond, Suzuki, Ryu, Abe,
  Barry, Bhattacharya, Donachie, Fukui, {et~al.}}]{bennett:2018a}
Bennett, D.~P., Udalski, A., Bond, I.~A., {et~al.} 2018, The Astronomical
  Journal, 156, 113

\bibitem[{{Bennett} {et~al.}(2020){Bennett}, {Bhattacharya}, {Beaulieu},
  {Blackman}, {Vand orou}, {Terry}, {Cole}, {Henderson}, {Koshimoto}, {Lu},
  {Baptiste Marquette}, {Ranc}, \& {Udalski}}]{bennett:2020a}
{Bennett}, D.~P., {Bhattacharya}, A., {Beaulieu}, J.-P., {et~al.} 2020, \aj,
  159, 68

\bibitem[{Bertin(2010)}]{bertin:2010a}
Bertin, E. 2010, Astrophysics Source Code Library

\bibitem[{{Bertin} \& {Arnouts}(1996)}]{bertin:1996a}
{Bertin}, E., \& {Arnouts}, S. 1996, \aaps, 117, 393

\bibitem[{{Bhattacharya} {et~al.}(2018){Bhattacharya}, {Beaulieu}, {Bennett},
  {Anderson}, {Koshimoto}, {Lu}, {Batista}, {Blackman}, {Bond}, {Fukui},
  {Henderson}, {Hirao}, {Marquette}, {Mroz}, {Ranc}, \&
  {Udalski}}]{bhattacharya:2018a}
{Bhattacharya}, A., {Beaulieu}, J.~P., {Bennett}, D.~P., {et~al.} 2018, \aj,
  156, 289

\bibitem[{Bhattacharya {et~al.}(2021)Bhattacharya, Bennett, Beaulieu, Bond,
  Koshimoto, Lu, Blackman, Vandorou, Terry, Batista,
  {et~al.}}]{bhattacharya:2021a}
Bhattacharya, A., Bennett, D.~P., Beaulieu, J.~P., {et~al.} 2021, AJ, 162, 60

\bibitem[{Bhattacharya {et~al.}(in prep)}]{bhattacharya:inprep}
Bhattacharya, A., {et~al.} in prep, {}

\bibitem[{Blackman {et~al.}(2021)Blackman, Beaulieu, Bennett, Danielski, Alard,
  Cole, Vandorou, Ranc, Terry, Bhattacharya, {et~al.}}]{blackman:2021a}
Blackman, J., Beaulieu, J., Bennett, D., {et~al.} 2021, Nature, 598, 272

\bibitem[{{Bond} {et~al.}(2001){Bond}, {Abe}, {Dodd}, {Hearnshaw}, {Honda},
  {Jugaku}, {Kilmartin}, {Marles}, {Masuda}, {Matsubara}, {Muraki}, {Nakamura},
  {Nankivell}, {Noda}, {Noguchi}, {Ohnishi}, {Rattenbury}, {Reid}, {Saito},
  {Sato}, {Sekiguchi}, {Skuljan}, {Sullivan}, {Sumi}, {Takeuti}, {Watase},
  {Wilkinson}, {Yamada}, {Yanagisawa}, \& {Yock}}]{bond01}
{Bond}, I.~A., {Abe}, F., {Dodd}, R.~J., {et~al.} 2001, \mnras, 327, 868

\bibitem[{{Bond} {et~al.}(2017){Bond}, {Bennett}, {Sumi}, {Udalski}, {Suzuki},
  {Rattenbury}, {Bozza}, {Koshimoto}, {Abe}, {Asakura}, {Barry},
  {Bhattacharya}, {Donachie}, {Evans}, {Fukui}, {Hirao}, {Itow}, {Li}, {Ling},
  {Masuda}, {Matsubara}, {Muraki}, {Nagakane}, {Ohnishi}, {Ranc}, {Saito},
  {Sharan}, {Sullivan}, {Tristram}, {Yamada}, {Yamada}, {Yonehara}, {Skowron},
  {Szyma{\'n}ski}, {Poleski}, {Mr{\'o}z}, {Soszy{\'n}ski}, {Pietrukowicz},
  {Koz{\l}owski}, {Ulaczyk}, \& {Pawlak}}]{bond17}
{Bond}, I.~A., {Bennett}, D.~P., {Sumi}, T., {et~al.} 2017, \mnras, 469, 2434

\bibitem[{{Boyajian} {et~al.}(2014){Boyajian}, {van Belle}, \& {von
  Braun}}]{boyajian:2014a}
{Boyajian}, T.~S., {van Belle}, G., \& {von Braun}, K. 2014, \aj, 147, 47

\bibitem[{Burrows(1987)}]{burrows:1987a}
Burrows, A. 1987, The Astrophysical Journal, 318, L57

\bibitem[{Choi {et~al.}(2012)Choi, Shin, Han, Udalski, Sumi, Gould, Bozza,
  Dominik, Fouque, Horne, {et~al.}}]{choi:2012a}
Choi, J.-Y., Shin, I.-G., Han, C., {et~al.} 2012, ApJ, 756, 48

\bibitem[{{Delfosse} {et~al.}(2000){Delfosse}, {Forveille}, {S{\'e}gransan},
  {Beuzit}, {Udry}, {Perrier}, \& {Mayor}}]{delfosse:2000a}
{Delfosse}, X., {Forveille}, T., {S{\'e}gransan}, D., {et~al.} 2000, \aap, 364,
  217

\bibitem[{Dong {et~al.}(2009)Dong, Bond, Gould, Koz{\l}owski, Miyake, Gaudi,
  Bennett, Abe, Gilmore, Fukui, {et~al.}}]{dong:2009a}
Dong, S., Bond, I., Gould, A., {et~al.} 2009, The Astrophysical Journal, 698,
  1826

\bibitem[{{Dong} {et~al.}(2009){Dong}, {Gould}, {Udalski}, {Anderson},
  {Christie}, {Gaudi}, {OGLE Collaboration}, {Jaroszy{\'n}ski}, {Kubiak},
  {Szyma{\'n}ski}, {Pietrzy{\'n}ski}, {Soszy{\'n}ski}, {Szewczyk}, {Ulaczyk},
  {Wyrzykowski}, {{\ensuremath{\mu}}FUN Collaboration}, {DePoy}, {Fox},
  {Gal-Yam}, {Han}, {L{\'e}pine}, {McCormick}, {Ofek}, {Park}, {Pogge}, {MOA
  Collaboration}, {Abe}, {Bennett}, {Bond}, {Britton}, {Gilmore}, {Hearnshaw},
  {Itow}, {Kamiya}, {Kilmartin}, {Korpela}, {Masuda}, {Matsubara}, {Motomura},
  {Muraki}, {Nakamura}, {Ohnishi}, {Okada}, {Rattenbury}, {Saito}, {Sako},
  {Sasaki}, {Sullivan}, {Sumi}, {Tristram}, {Yanagisawa}, {Yock}, {Yoshoika},
  {PLANET/RoboNet Collaborations}, {Albrow}, {Beaulieu}, {Brillant}, {Calitz},
  {Cassan}, {Cook}, {Coutures}, {Dieters}, {Dominis Prester}, {Donatowicz},
  {Fouqu{\'e}}, {Greenhill}, {Hill}, {Hoffman}, {Horne}, {J{\o}rgensen},
  {Kane}, {Kubas}, {Marquette}, {Martin}, {Meintjes}, {Menzies}, {Pollard},
  {Sahu}, {Vinter}, {Wambsganss}, {Williams}, {Bode}, {Bramich}, {Burgdorf},
  {Snodgrass}, {Steele}, {Doublier}, \& {Foellmi}}]{dong:2009b}
{Dong}, S., {Gould}, A., {Udalski}, A., {et~al.} 2009, \apj, 695, 970

\bibitem[{Freeman {et~al.}(in prep)}]{freeman:inprep}
Freeman, M., {et~al.} in prep, {}

\bibitem[{Fukui {et~al.}(2015)Fukui, Gould, Sumi, Bennett, Bond, Han, Suzuki,
  Beaulieu, Batista, Udalski, {et~al.}}]{fukui:2015a}
Fukui, A., Gould, A., Sumi, T., {et~al.} 2015, The Astrophysical Journal, 809,
  74

\bibitem[{Gaudi(1998)}]{gaudi:1998a}
Gaudi, B.~S. 1998, The Astrophysical Journal, 506, 533

\bibitem[{{Gaudi}(2012)}]{gaudi:2012a}
{Gaudi}, B.~S. 2012, \araa, 50, 411

\bibitem[{Ghez {et~al.}(2008)Ghez, Salim, Weinberg, Lu, Do, Dunn, Matthews,
  Morris, Yelda, Becklin, {et~al.}}]{ghez:2008a}
Ghez, A.~M., Salim, S., Weinberg, N., {et~al.} 2008, ApJ, 689, 1044

\bibitem[{{Gonzalez} {et~al.}(2011){Gonzalez}, {Rejkuba}, {Minniti}, {Zoccali},
  {Valenti}, \& {Saito}}]{gonzalez:2011a}
{Gonzalez}, O.~A., {Rejkuba}, M., {Minniti}, D., {et~al.} 2011, \aap, 534, L14

\bibitem[{Gould {et~al.}(2006)Gould, Udalski, An, Bennett, Zhou, Dong,
  Rattenbury, Gaudi, Yock, Bond, {et~al.}}]{gould:2006a}
Gould, A., Udalski, A., An, D., {et~al.} 2006, The Astrophysical Journal, 644,
  L37

\bibitem[{Han \& Gaudi(2008)}]{han:2008a}
Han, C., \& Gaudi, B.~S. 2008, The Astrophysical Journal, 689, 53

\bibitem[{{Henry} {et~al.}(1999){Henry}, {Franz}, {Wasserman}, {Benedict},
  {Shelus}, {Ianna}, {Kirkpatrick}, \& {McCarthy}}]{henry:1999a}
{Henry}, T.~J., {Franz}, O.~G., {Wasserman}, L.~H., {et~al.} 1999, \apj, 512,
  864

\bibitem[{{Henry} \& {McCarthy}(1993)}]{henry:1993a}
{Henry}, T.~J., \& {McCarthy}, Donald~W., J. 1993, \aj, 106, 773

\bibitem[{Hillman {et~al.}(2016)Hillman, Prialnik, Kovetz, \&
  Shara}]{hillman:2016a}
Hillman, Y., Prialnik, D., Kovetz, A., \& Shara, M.~M. 2016, The Astrophysical
  Journal, 819, 168

\bibitem[{Hunter(2007)}]{hunter:2007a}
Hunter, J.~D. 2007, CSE, 9, 90

\bibitem[{Janczak {et~al.}(2010)Janczak, Fukui, Dong, Monard, Koz{\l}owski,
  Gould, Beaulieu, Kubas, Marquette, Sumi, {et~al.}}]{janczak:2010a}
Janczak, J., Fukui, A., Dong, S., {et~al.} 2010, The Astrophysical Journal,
  711, 731

\bibitem[{{Kenyon} \& {Hartmann}(1995)}]{kenyon:1995a}
{Kenyon}, S.~J., \& {Hartmann}, L. 1995, \apjs, 101, 117

\bibitem[{Koshimoto {et~al.}(2021{\natexlab{a}})Koshimoto, Baba, \&
  Bennett}]{koshimoto:2021a}
Koshimoto, N., Baba, J., \& Bennett, D.~P. 2021{\natexlab{a}}, The
  Astrophysical Journal, 917, 78

\bibitem[{Koshimoto {et~al.}(2021{\natexlab{b}})Koshimoto, Bennett, Suzuki, \&
  Bond}]{koshimoto:2021b}
Koshimoto, N., Bennett, D.~P., Suzuki, D., \& Bond, I.~A. 2021{\natexlab{b}},
  ApJL, 918, L8

\bibitem[{Koshimoto \& Ranc(2021)}]{koshimoto:code}
Koshimoto, N., \& Ranc, C. 2021, {}, doi:10.5281/zenodo.4898012.
\newblock \url{https://doi.org/10.5281/zenodo.4898012}

\bibitem[{Lam {et~al.}(2022)Lam, Lu, Udalski, Bond, Bennett, Skowron, Mroz,
  Poleski, Sumi, Szymanski, {et~al.}}]{lam:2022a}
Lam, C.~Y., Lu, J.~R., Udalski, A., {et~al.} 2022, arXiv preprint
  arXiv:2202.01903

\bibitem[{Lu(2022)}]{lu:2022a}
Lu, J. 2022, {}, doi:10.5281/zenodo.6522913

\bibitem[{Lu {et~al.}(2008)Lu, Ghez, Hornstein, Morris, Becklin, \&
  Matthews}]{lu:2008a}
Lu, J., Ghez, A., Hornstein, S.~D., {et~al.} 2008, ApJ, 690, 1463

\bibitem[{{Minniti} {et~al.}(2010){Minniti}, {Lucas}, {Emerson}, {Saito},
  {Hempel}, {Pietrukowicz}, {Ahumada}, {Alonso}, {Alonso-Garcia}, {Arias},
  {Bandyopadhyay}, {Barb{\'a}}, {Barbuy}, {Bedin}, {Bica}, {Borissova},
  {Bronfman}, {Carraro}, {Catelan}, {Clari{\'a}}, {Cross}, {de Grijs},
  {D{\'e}k{\'a}ny}, {Drew}, {Fari{\~n}a}, {Feinstein}, {Fern{\'a}ndez
  Laj{\'u}s}, {Gamen}, {Geisler}, {Gieren}, {Goldman}, {Gonzalez}, {Gunthardt},
  {Gurovich}, {Hambly}, {Irwin}, {Ivanov}, {Jord{\'a}n}, {Kerins}, {Kinemuchi},
  {Kurtev}, {L{\'o}pez-Corredoira}, {Maccarone}, {Masetti}, {Merlo},
  {Messineo}, {Mirabel}, {Monaco}, {Morelli}, {Padilla}, {Palma}, {Parisi},
  {Pignata}, {Rejkuba}, {Roman-Lopes}, {Sale}, {Schreiber}, {Schr{\"o}der},
  {Smith}, {}, {Soto}, {Tamura}, {Tappert}, {Thompson}, {Toledo}, {Zoccali}, \&
  {Pietrzynski}}]{minniti:2010a}
{Minniti}, D., {Lucas}, P.~W., {Emerson}, J.~P., {et~al.} 2010, \na, 15, 433

\bibitem[{Nagakane {et~al.}(2017)Nagakane, Sumi, Koshimoto, Bennett, Bond,
  Rattenbury, Suzuki, Abe, Asakura, Barry, {et~al.}}]{nagakane:2017a}
Nagakane, M., Sumi, T., Koshimoto, N., {et~al.} 2017, The Astronomical Journal,
  154, 35

\bibitem[{{Nataf} {et~al.}(2013){Nataf}, {Gould}, {Fouqu{\'e}}, {Gonzalez},
  {Johnson}, {Skowron}, {Udalski}, {Szyma{\'n}ski}, {Kubiak},
  {Pietrzy{\'n}ski}, {Soszy{\'n}ski}, {Ulaczyk}, {Wyrzykowski}, \&
  {Poleski}}]{nataf:2013a}
{Nataf}, D.~M., {Gould}, A., {Fouqu{\'e}}, P., {et~al.} 2013, \apj, 769, 88

\bibitem[{{Nishiyama} {et~al.}(2009){Nishiyama}, {Tamura}, {Hatano}, {Kato},
  {Tanab{\'e}}, {Sugitani}, \& {Nagata}}]{nishiyama:2009a}
{Nishiyama}, S., {Tamura}, M., {Hatano}, H., {et~al.} 2009, \apj, 696, 1407

\bibitem[{Oliphant(2006)}]{oliphant:2006a}
Oliphant, T.~E. 2006, A guide to NumPy, Vol.~1 (Trelgol Publishing USA)

\bibitem[{Revnivtsev {et~al.}(1998)Revnivtsev, Gilfanov, \&
  Churazov}]{revnivtsev:1998a}
Revnivtsev, M., Gilfanov, M., \& Churazov, E. 1998, Arxiv preprint
  astro-ph/9804173

\bibitem[{Sahu {et~al.}(2022)Sahu, Anderson, Casertano, Bond, Udalski, Dominik,
  Calamida, Bellini, Brown, Rejkuba, {et~al.}}]{sahu:2022a}
Sahu, K.~C., Anderson, J., Casertano, S., {et~al.} 2022, arXiv preprint
  arXiv:2201.13296

\bibitem[{{Service} {et~al.}(2016){Service}, {Lu}, {Campbell}, {Sitarski},
  {Ghez}, \& {Anderson}}]{service:2016a}
{Service}, M., {Lu}, J.~R., {Campbell}, R., {et~al.} 2016, \pasp, 128, 095004

\bibitem[{{Stetson}(1987)}]{stetson:1987a}
{Stetson}, P.~B. 1987, \pasp, 99, 191

\bibitem[{{Sumi} {et~al.}(2003){Sumi}, {Abe}, {Bond}, {Dodd}, {Hearnshaw},
  {Honda}, {Honma}, {Kan-ya}, {Kilmartin}, {Masuda}, {Matsubara}, {Muraki},
  {Nakamura}, {Nishi}, {Noda}, {Ohnishi}, {Petterson}, {Rattenbury}, {Reid},
  {Saito}, {Saito}, {Sato}, {Sekiguchi}, {Skuljan}, {Sullivan}, {Takeuti},
  {Tristram}, {Wilkinson}, {Yanagisawa}, \& {Yock}}]{sumi:2003a}
{Sumi}, T., {Abe}, F., {Bond}, I.~A., {et~al.} 2003, \apj, 591, 204

\bibitem[{Suzuki {et~al.}(2013)Suzuki, Udalski, Sumi, Bennett, Bond, Abe,
  Botzler, Freeman, Fukagawa, Fukui, {et~al.}}]{suzuki:2013a}
Suzuki, D., Udalski, A., Sumi, T., {et~al.} 2013, The Astrophysical Journal,
  780, 123

\bibitem[{{Suzuki} {et~al.}(2016){Suzuki}, {Bennett}, {Sumi}, {Bond}, {Rogers},
  {Abe}, {Asakura}, {Bhattacharya}, {Donachie}, {Freeman}, {Fukui}, {Hirao},
  {Itow}, {Koshimoto}, {Li}, {Ling}, {Masuda}, {Matsubara}, {Muraki},
  {Nagakane}, {Onishi}, {Oyokawa}, {Rattenbury}, {Saito}, {Sharan}, {Shibai},
  {Sullivan}, {Tristram}, {Yonehara}, \& {MOA Collaboration}}]{suzuki:2016a}
{Suzuki}, D., {Bennett}, D.~P., {Sumi}, T., {et~al.} 2016, \apj, 833, 145

\bibitem[{{Suzuki} {et~al.}(2018){Suzuki}, {Bennett}, {Ida}, {Mordasini},
  {Bhattacharya}, {Bond}, {Donachie}, {Fukui}, {Hirao}, {Koshimoto},
  {Miyazaki}, {Nagakane}, {Ranc}, {Rattenbury}, {Sumi}, {Alibert}, \&
  {Lin}}]{suzuki:2018a}
{Suzuki}, D., {Bennett}, D.~P., {Ida}, S., {et~al.} 2018, \apjl, 869, L34

\bibitem[{{Szyma{\'n}ski} {et~al.}(2011){Szyma{\'n}ski}, {Udalski},
  {Soszy{\'n}ski}, {Kubiak}, {Pietrzy{\'n}ski}, {Poleski}, {Wyrzykowski}, \&
  {Ulaczyk}}]{ogle3-phot}
{Szyma{\'n}ski}, M.~K., {Udalski}, A., {Soszy{\'n}ski}, I., {et~al.} 2011,
  \actaa, 61, 83

\bibitem[{Terry {et~al.}(2021)Terry, Bhattacharya, Bennett, Beaulieu,
  Koshimoto, Blackman, Bond, Cole, Henderson, Lu, {et~al.}}]{terry:2021a}
Terry, S.~K., Bhattacharya, A., Bennett, D.~P., {et~al.} 2021, AJ, 161, 54

\bibitem[{Udalski {et~al.}(1993)Udalski, Szymanski, Kaluzny, Kubiak,
  Krzeminski, Mateo, Preston, \& Paczynski}]{udalski:1993a}
Udalski, A., Szymanski, M., Kaluzny, J., {et~al.} 1993, Acta astronomica, 43,
  289

\bibitem[{Udalski {et~al.}(2015)Udalski, Szyma{\'n}ski, \&
  Szyma{\'n}ski}]{udalski:2015a}
Udalski, A., Szyma{\'n}ski, M., \& Szyma{\'n}ski, G. 2015, arXiv preprint
  arXiv:1504.05966

\bibitem[{{Yelda} {et~al.}(2010){Yelda}, {Lu}, {Ghez}, {Clarkson}, {Anderson},
  {Do}, \& {Matthews}}]{yelda:2010a}
{Yelda}, S., {Lu}, J.~R., {Ghez}, A.~M., {et~al.} 2010, \apj, 725, 331

\bibitem[{{Zhang} {et~al.}(2022){Zhang}, {Gaudi}, \& {Bloom}}]{zhang:2022a}
{Zhang}, K., {Gaudi}, B.~S., \& {Bloom}, J.~S. 2022, Nature Astronomy,
  doi:10.1038/s41550-022-01671-6

\end{thebibliography}

\end{document}